\documentclass{emulateapj}
\usepackage{amsmath}
\usepackage{natbib}
\usepackage{float}
\usepackage{graphicx}
\usepackage{subfigure}
\usepackage{xcolor}
\usepackage[dvipdfm]{hyperref}

\def\msun{{M}_{\odot}}
\def\medd{{\dot M_{Edd}}}

\def\be{\begin{equation}}
\def\ee{\end{equation}}
\def\msunyr{\msun~{\rm yr}^{-1}}
\def\ergs{\rm erg~s^{-1}}
\def\Edotb{{\dot E_{b}}}
\def\Mdotw{{\dot M_{w}}}
\def\Mc{{M_{c}}}
\def\Rdot2{{\dot R_{2}}}

\catcode`\@=11 
\def\@versim#1#2{\vcenter{\offinterlineskip
        \ialign{$\m@th#1\hfil##\hfil$\crcr#2\crcr\sim\crcr } }}

\shorttitle{{\it Fermi} bubbles inflated by winds from hot accretion flow in Sgr A*}
\shortauthors{Mou et al}

\begin{document}
\bibliographystyle{apj}
\title
{{\it Fermi} Bubbles inflated by winds launched from the hot accretion flow in Sgr A*}

\author {Guobin Mou\altaffilmark{1,2}, Feng Yuan\altaffilmark{1}, Defu Bu\altaffilmark{1}, Mouyuan Sun\altaffilmark{3,4} , and Meng Su\altaffilmark{5,6}}

 \altaffiltext{1}{Shanghai Astronomical Observatory, Chinese Academy of Sciences, 80 Nandan Rd, Shanghai 200030, China; gbmou@shao.ac.cn, fyuan@shao.ac.cn}
 \altaffiltext{2}{University of Chinese Academy of Sciences, 19A Yuquan Road, Beijing 100049, China}
 \altaffiltext{3}{Department of Astronomy, and Institute of Theoretical Physics and Astrophysics, Xiamen University, Xiamen, Fujian 361005, China}
 \altaffiltext{4}{Department of Astronomy and Astrophysics, and Institute for Gravitation and the Cosmos, Pennsylvania State University, University Park, PA 16802, USA}
 \altaffiltext{5}{Department of Physics, and Kavli Institute for Astrophysics and Space Research, Massachusetts Institute of Technology, Cambridge, MA 02139, USA}
 \altaffiltext{6}{Einstein Fellow}

\begin{abstract}

A pair of giant gamma-ray bubbles have been revealed by
 the {\it Fermi} LAT. In this paper we investigate their formation mechanism. Observations have indicated that the activity of the supermassive black hole located at the Galactic center, Sgr A*, was much stronger than the present time. Specifically, one possibility is that while Sgr A* was also in the hot accretion regime, the accretion rate should be $10^3-10^4$ times higher during the past $\sim 10^7$ yr. On the other hand, recent MHD numerical simulations of hot accretion flows have unambiguously shown the existence of strong winds and obtained their properties. Based on these knowledge, by performing three-dimensional hydrodynamical simulations, we show in this paper that the Fermi bubbles could be inflated by winds launched from the ``past'' hot accretion flow in Sgr A*.  In our model, the active phase of Sgr A* is required to last for about 10 million years and it was quenched no more than 0.2 million years ago. The Central Molecular Zone (CMZ) is included and it collimates the wind orientation towards the Galactic poles. Viscosity suppresses the Rayleigh-Taylor and Kelvin-Helmholtz instabilities and results in the smoothness of the bubble edge. The main observational features of the bubbles can be well explained. Specifically, the {\it ROSAT} X-ray features are interpreted by the shocked interstellar medium and the interaction region between winds and CMZ gas. The thermal pressure and temperature obtained in our model are in good consistency with the recent {\it Suzaku} observations.

\end{abstract}


\section{INTRODUCTION}

 Observations have shown that there exists a supermassive black hole, Sgr A*, located at the Galactic Center (GC). The mass of the black hole is about $4~\times ~10^6 \msun$ (\citealt{Schodel2002}; \citealt{Ghez2005,Ghez2008}; \citealt{Gillessen2009a,Gillessen2009b}). Because of its proximity, Sgr A* is regarded as the best laboratory of studying black hole accretion. Numerous observations have been conducted and abundant data has been obtained (see recent reviews by \citealt{Genzel2010}; Falcke \& Markoff 2013; Yuan \& Narayan 2014). The source is quite dim currently, with a bolometric luminosity of only about $10^{36}~\ergs~\sim 3\times 10^{-9}~L_{Edd}$.
 The mass accretion rate at the Bondi radius has been estimated by combining the {\it Chandra} observation and the Bondi accretion theory, which is  $\sim 10^{-5} \msunyr$ (\citealt{Baganoff2003}). The bolometric luminosity would be 5 orders of magnitude higher if the accretion were in the mode of the standard thin disk. Amount of theoretical studies in the past 20 years have revealed that the advection-dominated accretion flow (ADAF) can explain this puzzle (\citealt{Yuan2003}). Specifically, the low-luminosity of Sgr A* is because of two reasons. One is the intrinsic low radiative efficiency of ADAF because of energy advection (\citealt{Narayan1994,Narayan1995b}; \citealt{Xie2012}).
 Another important reason is because of the existence of strong wind (or outflow), i.e., $\sim 99\%$ of the matter captured at the Bondi radius are lost (\citealt{Yuan2012b}; \citealt{Narayan2012}; \citealt{LiOS2013}). The existence of wind has been confirmed by the radio polarization observations (e.g., \citealt{Aitken2000}; \citealt{Bower2003}; \citealt{Marrone2007}), and more recently by the {\it Chandra} observation to the emission lines from the accretion flow in Sgr A* (\citealt{Wang2013}). Yuan \& Narayan (2014) presented the most recent review on the hot accretion flow and its various astrophysical applications, including on Sgr A*.

One particularly interesting thing is that many observational evidences show that the activity of Sgr A* was very likely much stronger in the past than the current stage. These observations suggest that Sgr A* has perhaps undergone multiple past epochs of enhanced activity on different timescales. Here we only focus on relatively long timescales. These evidences were summarized in \citet{Totani2006}, and later discussed in other works (e.g., \citealt{Bland-Hawthorn2013}; \citealt{Ponti2013}; \citealt{Kataoka2013}).
These evidences include: 1) orders of magnitude higher X-ray luminosity (compared to the present value) required to explain the fluorescent X-ray emission reflected from cold iron atoms in the giant molecular cloud Sgr B2 (\citealt{Koyama1996}; \citealt{Murakami_B22000,Murakami_B22001}; \citealt{Revnivtsev2004}); 2) a new X-ray reflection nebula associated with Sgr C detected by {\it ASCA} (\citealt{Murakami_C2001});
3) the ionized halo surrounding Sgr A* (Maeda et al. 2002); 4) Galactic Center Lobe (GCL, \citealt{Bland-Hawthorn2003}); 5) Expanding Molecular Ring (EMR, \citealt{Kaifu1972}; \citealt{Scoville1972}); 6) North Polar Spur (NPS, \citealt{Sofue2000}; \citealt{Bland-Hawthorn2003}); 7) the 8 keV diffuse X-ray emission in the center (\citealt{Muno2004}); 8) the excess of H$\alpha$ emission of Magellanic Stream (\citealt{Bland-Hawthorn2013}); 9) the {\it Suzaku} observations to the NPS (\citealt{Kataoka2013}). \citet{Totani2006} found that to explain the former seven observations mentioned above,
 the characteristic X-ray luminosity of Sgr A* should be $\sim (10^{39}-10^{40})~\ergs\sim 2\times (10^{-6}-10^{-5})~L_{\rm Edd}$ several hundred years ago, and such an activity should last for $\sim 10^7$ yr.
For such a luminosity, the accretion should be well in the regime of hot accretion rather than the standard thin disk (Yuan \& Narayan 2014). Correspondingly, the mass accretion rate should be $10^3-10^4$ times higher than the present value (\citealt{Totani2006}).
Other possibilities of the past activity have also been proposed. For example, the bolometric luminosity in the past millions years estimated by \citet{Bland-Hawthorn2013} based on the 8th evidence mentioned above is much higher, $\sim 0.03-0.3~L_{\rm Edd}$. The timescale of the activity is shorter, and it was active 1$-$3 Myr ago. Yet another possibility is as follows. A star formation event has been observed and it is believed to occur at $\sim 6\times 10^6$ yr ago on scales of $\sim$ 0.03$-$0.5 pc from the SMBH (e.g., \citealt{Genzel2003}; \citealt{Paumard2006}). If the past activity of Sgr A* occur concurrently with this event, this would imply a strong activity of Sgr A* occurred $\sim 6$ Myr ago (\citealt{Zubovas2011}). In summary, so far we still lack a consensus on the past activity of Sgr A*.

Yet perhaps another evidence for the past activity of Sgr A* is the {\it Fermi} bubbles recently detected. Using the {\it Fermi}-LAT, \citet{Su10} discovered two giant gamma-ray bubbles located above and below the Galactic plane
(also refer to \citealt{YangRZ2014} for the recent observations). In Galactic coordinates (l,b), the height of each bubble is about $50^{\circ}$, and the width is about $40^{\circ}$. The surface brightness looks uniform, and the edge looks sharp. The total luminosity of the bubbles is $4 \times 10^{37}~\ergs$ in 1$-$100 GeV band. The total energy of the two bubbles is estimated to be $10^{55}-10^{56}$ erg.

Many theoretical models have been proposed since the discovery of the {\it Fermi} Bubbles.
In the ``hadronic'' model, the formation is explained as due to a population of relic cosmic ray protons injected by processes associated with extremely long time scale and high areal density star formation in the Galactic center (\citealt{Crocker2011,Crocker2012,Crocker2013}).
In the ``leptonic'' scenario the $\gamma$-ray emission comes from the inverse Compton scattering between relativistic electrons (also often called as Cosmic Ray) and seed photons. The seed photons may be the cosmic microwave background, but the origin of relativistic electrons are different in different models. They can come from Fermi-1st order acceleration on shock front formed in the periodic star capture processes by Sgr A* (\citealt{Cheng2011}), the Fermi-2nd order acceleration through stochastic scattering by plasma instabilities (\citealt{Mertsch2011}), directly from the jet (\citealt{Guo1}; \citealt{Guo2}; \citealt{Yang2012}; \citealt{Yang2013}), or from outflows driven by the past star formation (\citealt{Carretti2013}).

Among these models, there are two models which are physically most relevant to the model we  propose in the present paper. They are the ``jet'' model (\citealt{Guo1}; \citealt{Guo2}) and the ``quasar outflow'' model (\citealt{Zubovas2011}; \citealt{Zubovas2012}). In the former, it is suggested that the bubbles are created by AGN jet which happened about 2 Myr ago. After that, cosmic rays (CRs) carried by jet diffuse to today's morphology. Yang et al. (2012, 2013) developed the jet model by including magnetic field. They showed that the suppression of the diffusion of CRs along the direction across the edge is caused by the magnetic field configuration. This is because inside the bubbles, the magnetic field is mainly radial, but just outside of the bubble and close to the edge, the field is mainly in the parallel direction. One problem, as pointed out by \citet{Zubovas2011}, is that they must require the jet direction to be perpendicular to the plane of the Galaxy, which seems to be unlikely, given the general absence of correlation between the direction of jets and galaxy planes and the observed direction of the stellar disk in the Galaxy. In addition, the velocity required in the jet model is as low as $\leq0.1c$ and the mass loss rate in the jet is in general as high as super-Eddington.

Another model is the ``quasar outflow'' model proposed in \citet{Zubovas2011} and \citet{Zubovas2012}. In this model, Sgr A* is again assumed to be very active in the past, with mildly super-Eddington accretion rate 6 Myr ago and duration of the activity being 1 Myr. Under such a high luminosity, quasi-spherical outflow will be driven by the strong radiation pressure from this quasar (\citealt{KingPounds2003}), which can result in the formation of the {\it Fermi} bubbles. In this model, the existence of the well-known central molecular zone (CMZ) in the GC region plays an important role in collimating the outflow and forming the morphology of the bubbles. \citet{Kataoka2013} pointed out that the expansion velocity derived by the {\it Suzaku} observation is lower than the advocated values by both the jet and quasar outflow models
 by a factor of 5 and 2 respectively.

Assuming that Sgr A* was in an active state as suggested by Totani (2006), in this paper we investigate whether the {\it Fermi} bubbles can be inflated by the wind launched from the hot accretion flow by performing numerical simulations. In \S2, we briefly introduce some background on the accretion flow and wind, and present an analytical solution for the interaction between the winds and ISM to be used to understand our numerical simulation results. The numerical simulations approach and the results are presented in \S3 and \S4, respectively. We then summarize in \S5.

\section{Models}
\subsection{Accretion flows in Sgr A*}

The accretion flow in Sgr A* in the current stage is relatively simple, namely the whole accretion flow is hot, ranging from the Bondi radius to the black hole horizon. However, if the accretion rate is $10^3-10^4$ higher, as estimated by Totani (2006), this simple picture needs to be modified. Numerous observational and theoretical studies have shown that the accretion flow should consist of an outer thin disk and an inner hot accretion flow. The boundary between the truncated thin disk and the hot accretion flow is called the transition radius ($R_{\rm tr}$). Some works have been done on the physical mechanism of the transition. Although this question is still not completely solved, we now have a consensus that the value of $R_{\rm tr}$ should decrease with increasing mass accretion rate. This is supported by the modeling to some low-luminosity AGNs and the hard state of black hole X-ray binaries, which is summarized in \citet{YuanNarayan2004}.

For hot accretion flow, the mass accretion rate is a function of radius because of the mass loss in the wind throughout the disk (refer to \S\ref{wind}). The current net mass accretion rate at the horizon of the black hole and at the Bondi radius are $\sim 10^{-7}\dot{M}_{\rm Edd}$ and $10^{-5}\dot{M}_{\rm Edd}$, respectively (\citealt{Yuan2003}). Here $\dot{M}_{\rm Edd}\equiv 10L_{\rm Edd}/c^2$ is defined as the Eddington accretion rate. According to \citet{Totani2006}, the mass accretion rate close to the horizon of black hole in Sgr A* should be $10^{-4}-10^{-3}~\dot{M}_{\rm Edd}$ during the past $10^7$ yr. For this value of accretion rate, given the theoretical uncertainty, $R_{\rm tr}=500 R_s$ would be a reasonable assumption, here $R_s=2GM/c^2$ is the Schwarzschild radius of the black hole. Note that there is some uncertainty in the value of $R_{\rm tr}$. The mass accretion rate at $R_{\rm tr}=500 R_s$ is set to be \begin{equation} \dot{M}_{\rm acc}(500R_s)\approx 0.02\dot{M}_{\rm Edd}\end{equation} in our favored model. This value is $2\times 10^3$ times higher than the present value, well within the range obtained in \citet{Totani2006}.

\subsection{Wind}
\label{wind}

As we have mentioned in \S1, one characteristic feature of hot accretion flow is that it is subject to strong wind. The existence of wind has been suggested in \citet{Narayan1994} and later by \citet{BlandfordBegelman1999}. The hydrodynamic (HD) and magnetohydrodynamic (MHD) numerical simulation works by \citet{Stone99} and \citet{Stone2001} showed that the mass inflow rate of the accretion flow decreases inward, which can be regarded as the pioneer works in the quantitative study of winds from hot accretion flow.
This result is confirmed by many other subsequent works (see review by \citealt{Yuan2012a}). It was soon shown that the physical reason for the inward decrease of inflow rate is because of mass loss in wind which occurs in a wide range of radius throughout the accretion flow (\citealt{Yuan2012b}; \citealt{Narayan2012}; \citealt{LiOS2013}; \citealt{Sadowski2013}). The physical mechanism for the production of winds is found to be the combination of magnetocentrifugal force and the gradient of gas and magnetic pressure (\citealt{Yuan2012b}; Yuan et al. in preparation). While the existence of wind is evident, consensus on some quantitative features of the wind have not been reached. For example, \citet{Yuan2012b} argued that the mass flux of wind should be significant, comparable to the mass flux of inflow. This is much larger than the lower limit obtained in \citet{Narayan2012}. In this work, we follow \citet{Yuan2012b} and assume that at $R_{\rm tr}$, the mass flux of wind is roughly equal to the inflow rate there, i.e.,
\begin{equation} \dot{M}_{\rm wind}\approx 0.02\dot{M}_{\rm Edd} \end{equation} in most of our models except for runs ``G'' and ``H'' (refer to Table 1).

\citet{Yuan2012b} (see also \citealt{LiOS2013}) also estimated the terminal radial velocity of wind based on the conservation of the value of Bernoulli parameter $Be$ and found it is roughly half of the Keplerian velocity at $R_{\rm tr}$. However, that estimation should be regarded as the lower limit since magnetic field is not included in the analysis. Our more recent study find that $Be$ actually increases along the streamline when magnetic field is included (Yuan et al. 2014, in preparation). In the present work we set the velocity of the wind to be \begin{equation} v_{\rm wind}\approx 2v_{\rm k}(500 R_s).\end{equation} Our simulations indicate that there is some degeneracy between the mass flux and the velocity of winds. What really matters is the power of winds. The mass flux and velocity adopted above correspond to the power of wind $P_{w}=2\times 10^{41}~\ergs$.

The next wind parameter is their angular distribution. In spherical coordinate, Yuan et al. (in preparation) find that winds occupy a region $\theta\sim0^{\circ}-60^{\circ}$ and $\theta \sim 120^{\circ}-180^{\circ}$. Given that the range is quite large, combining with the possibility that during the long timescale of $10^7$ yr the rotation axis of the accretion flow may have changed with time, in the present work, we simply assume that the winds are blown out isotropically.

Winds may also be launched from the truncated thin disk outside of $R_{\rm tr}$. But the details of this process is poorly investigated at present. In this work, we assume that this part of wind is not important compared to the winds from the inner hot accretion flow. This is the main uncertainty of our model.

\subsection{Shock}
The winds launched from the hot accretion flow are usually supersonic so they will interact with the interstellar medium and produce shocks. Before we present the details of our simulation results, in this part we present some analytical solutions to this problem based on some simplifications, which is helpful to understanding our simulation results.
Here we assume a simple shock model formed by an isotropic wind punching into an isotropic distribution of
interstellar medium (ISM). It is well-known that the region can be divided into the following four parts: 1) high speed wind; 2) shocked wind; 3) shocked ISM; 4) un-shocked ISM gas. The interface between the shocked wind and shocked ISM is called the contact discontinuity (CD). In our case, since the cooling timescale of shocked winds is longer than the flow timescale, the shocked wind is a ``energy-driven'' flow rather than ``momentum-driven'' flow (\citealt{King2003}, \citealt{Zubovas2011}, \citealt{ZubovasKing2012} and \citealt{Faucher-Giguere2012}). If we assume that the forward shock velocity $\dot R_{2}$ is equal to the velocity of the shocked ISM ($v_{\bf c}$) and the shocked ISM region is so thin that $R_{c} \sim R_{2}$, approximately we can obtain the following equations (e.g., \citealt{Castor1975}, Weaver et al. 1977):
  \begin{gather} 
    \Edotb=P_{w}-4\pi R^{2}_{2} P_{b} {\dot R_{2}},  \\ 
    E_{b}\approx \frac{4}{3}\pi R^{3}_{2} \cdot \frac{3}{2} P_{b},  \\ 
    \frac{d}{dt}(\Mc \Rdot2)=4\pi R^{2}_2 P_{b},  \\ 
    \Mc=\int^{R_{2}}_{0} \rho_{\texttt{ISM}} 4\pi r^{2}dr,  \\ 
    P_{w}=\frac{1}{2}\Mdotw V^{2}.
  \end{gather}  
Here, $E_{b}$ is the total energy of shocked wind, in which the internal energy is dominant, $P_{w}$ is the kinetic power of the un-shocked wind, $P_{b}$ is the gas pressure of shocked wind. In the case of weak shock, the shock velocity $\dot R_{2}$ will be significantly higher than the velocity of the shocked ISM, so our approximations may introduce large errors. Assuming $\rho_{\texttt{ISM}}=Ar^{-n}$, in which \emph{A} and \emph{n} are both constants and $n<3$,  the solutions of the above equations are:
 \begin{gather} 
  R_{2}(t)=f(A,n)\cdot P_{w}^{\frac{1}{5-n}} t^{\frac{3}{5-n}},  \label{Equ_R2} \\ 
  P_{b}(t)=g(A,n)\cdot P_{w}^{\frac{2-n}{5-n}} t^{-\frac{n+4}{5-n}}, \label{Equ_P} \\ 
  f(A,n)=\left[\frac{(5-n)^{3}(3-n)}{14\pi A(7-2n)(11-n)}\right]^{\frac{1}{5-n}}, \label{Equ_Rfan} \\
  g(A,n)=\frac{3A(7-2n)}{(5-n)^{2}(3-n)}\cdot f(A,n)^{2-n}.
 \end{gather}
 Here, $R_{2}$ is the radii of the forward shock, and roughly it can be used to represent the radii of CD. Here we have neglected the gravity. This is because in our case the work done by overcoming the gravity is one order of magnitude lower than the injected energy from Sgr A*. Our solution of $R_{2}$ is similar to the energy-driven solution in \citet{Faucher-Giguere2012}. Besides, we find that when the density profile is assumed to be the same as \citet{Zubovas2011}, the velocity of the shocked ISM is also close to their result.

\begin{table*}
  \centering
  \begin{minipage}{130mm}
  \renewcommand{\thefootnote}{\thempfootnote}
  \caption{Parameters of simulations}
  \begin{tabular}{@{} c  c  c  c  c  c  c  c  c  c }
    \hline
        & $n_{e0}$
        & $R_{t}$
        & {$v_{j}$ \footnote{$v_{j}$ is the velocity of the wind.}}
        & $\mu$
        & \rm $\dot {M}_{out}$
        & {$P_{w}$\footnote{$P_{w}$ is the total kinetic power of the wind injected in $4\pi$ of solid angle.}}
        & {$t_{\rm FB}$ \footnote{$t_{\rm FB}$ is the age of {\it Fermi} Bubbles.}}
        & {$t_{\rm Q}$ \footnote{$t_{\rm Q}$ is the duration of the quiescent state of Sgr A* in the final stage.}} \\
    Run & $(\text{cm}^{-3}$) & ($\rm Rs$) & (c) & ($\rm g~cm^{-1}~{s}^{-1}$) & ($\medd$) & ($10^{41}~\rm erg~s^{-1} $) & (\rm  Myr) & (\rm  Myr)\\
    \hline

    A & $1.0\times 10^{-2}$ & $5\times10^{2} $ & 6.2\% & 2.0 & 2.0\% & 2.0 & 12.3 & - \\ 
    B & $1.0\times 10^{-2}$ & $5\times10^{2} $ & 6.2\% & 2.0 & 2.0\% & 2.0 & 12.3 & 0.3 \\ 
    C & $1.0\times 10^{-2}$ & $5\times10^{2} $ & 6.2\% &  0  & 2.0\% & 2.0 &  7.6 & - \\ 
    D & $1.0\times 10^{-2}$ & $5\times10^{2} $ & 6.2\% & 4.0 & 2.0\% & 2.0 & 14.5 & - \\ 
    E\footnote{The only difference between E and A is that thermal conductivity is not considered in E.} &$1.0\times 10^{-2}$ & $5\times10^{2} $ & 6.2\% & 2.0 & 2.0\% & 2.0 & 13.4 & -  \\  
    F & $2.0\times 10^{-2}$ & $5\times10^{2} $ & 6.2\% & 2.0 & 2.0\% & 2.0 & 14.9 & - \\ 
    G & $1.0\times 10^{-2}$ & $5\times10^{2} $ & 6.2\% & 3.0 & 6.0\% & 6.0 &  8.1 & - \\ 
    H & $1.0\times 10^{-2}$ & $1\times10^{3} $ & 4.3\% & 2.0 & 4.2\% & 2.0 & 11.6 & - \\ 
    \hline
 \label{table1}
  \end{tabular}
 \end{minipage}
\end{table*}

\begin{table*}
 \centering
  \begin{minipage}{130mm}
  \renewcommand{\thefootnote}{\thempfootnote}
  \caption{Results}
  \begin{tabular}{@{}  c  c  c  c  c  c  c  c  c  c  c }
    \hline
    & {H/W \footnote{Height/Width of the {\it Fermi} Bubble, in unit of kpc/kpc}}
    & {$T_{\rm FB}$\footnote{$T_{\rm FB}$ is the space-averaging temperature of {\it Fermi} Bubbles}}
    & {$T_{\rm X}$\footnote{$T_{\rm X}$ is the space-averaging temperature of shocked ISM.}}
    & {$E_{\rm FB}$\footnote{$E_{\rm FB}$ is the internal energy of {\it Fermi} Bubbles.}}
    & {$K_{\rm FB}$\footnote{$K_{\rm FB}$ is the kinetic energy of {\it Fermi} Bubbles.}}
    & {$E_{\rm X}$\footnote{$E_{\rm X}$ is the internal energy of shocked ISM.}}
    & {$K_{\rm X}$\footnote{$K_{\rm X}$ is the kinetic energy of shocked ISM.}}
    & {$E_{\rm inj}$\footnote{$E_{\rm tot}$ is the total energy injected by Sgr A* wind.}}
    & {$M_{inj}$\footnote{$M_{inj}$ is the total mass injected from the origin.}}
    & {$M_{\rm FB}$\footnote{$M_{\rm FB}$ is the total mass of {\it Fermi} Bubbles.} }  \\
    Run &   & ($10^{8}$ \rm{K}) & ($10^{6}$ \rm{K}) & ($10^{55}~\rm{erg}$) & ($10^{55}~\rm{erg}$) & ($10^{55}~\rm{erg}$) & ($10^{55}~\rm{erg}$) & ($10^{55}~\rm{erg}$) & ($\msun$) & ($\msun$) \\
    \hline

    A & 8/7  & 5 & 3 & 2.2 & 0.2 & 3.5 & 2.1 & 7.7 & $2\times10^{4}$ & $2.5\times10^{5}$ \\ 
    B & 8/7  & 5 & 3 & 2.2 & 0.1 & 3.6 & 2.1 & 7.6 & $2\times10^{4}$ & $2.5\times10^{5}$ \\ 
    C & 10/4 & 3 & 3 & 0.7 & 0.4 & 1.9 & 0.9 & 4.7 & $1\times10^{4}$ & $1.2\times10^{5}$ \\ 
    D & 8/8.5& 5 & 3 & 2.7 & 0.2 & 4.4 & 2.8 & 9.2 & $3\times10^{4}$ & $2.9\times10^{5}$ \\ 
    E & 8/8  & 10& 3 & 2.5 & 0.2 & 3.9 & 2.5 & 8.4 & $2\times10^{4}$ & $3.0\times10^{5}$ \\ 
    F & 8/7  & 5 & 3 & 2.7 & 0.2 & 5.2 & 2.4 & 9.4 & $3\times10^{4}$ & $3.5\times10^{5}$ \\ 
    G & 8/6.5& 7 & 5 & 4.5 & 0.7 & 5.7 & 4.6 & 15.2& $4\times10^{4}$ & $4.5\times10^{5}$ \\ 
    H & 8/7  & 4 & 3 & 2.0 & 0.3 & 3.3 & 2.0 & 7.2 & $4\times10^{4}$ & $2.8\times10^{5}$ \\ 
    \hline
 \label{table2}
  \end{tabular}
 \end{minipage}
\end{table*}

\section{SIMULATION}
 \subsection{Simulation Setup}
 We use ZEUS code (\citealt{Stone92}; \citealt{Hayes2006}) and adopt 3-D Cartesian coordinates. The advantage of choosing Cartesian coordinates rather than spherical or cylindrical coordinates is  that we can avoid the singularity on the polar axis arisen by one term of the viscous stress tensor. Computational domain is from $-$6.4 kpc to +6.4 kpc in the X-,Y-direction, and 0$-$12 kpc in the Z-direction. Z-axis stretches along the Galactic pole, and X-Y plane is the Galactic plane. Sgr A* is located at the origin. We adopt non-uniform grid, with $\bigtriangleup x_{i+1}/\bigtriangleup x_{i}=1.062$, $\bigtriangleup y_{j+1}/\bigtriangleup y_{j}=1.062$, and $\bigtriangleup z_{k+1}/\bigtriangleup z_{k}=1.035$. The numbers of meshes are I=128, J=128 and K=120 in X-,Y-, and Z-direction respectively. We use the reflecting boundary condition on the lower boundary ($Z=0$), and choose the outflow boundary condition on the other five boundary surface.

\subsection{Initial Conditions}
We assume that the initial interstellar medium is an isothermal sphere in hydrostatic equilibrium state, i.e., the gradient of the gas pressure balances the gravity. Specifically, we assume the gravitational force given by stars and dark matter in a simplified form:
 \be \nabla\phi(r)=-\frac{2\sigma^{2}}{r}\vec{r}, \ee where $\vec{r}=\vec{x}+\vec{y}+\vec{z}$.
It will give a constant velocity dispersion of stars, and the velocity dispersion is 100 ${\rm km~s^{-1}}$ here. This is very similar to the circumstance in the galactic bulge. In the recent work by \citet{Miller2013}, a $\beta$-model was assumed to describe the gas density profile of Galactic hole, and $n_{e}$ scales from $10^{-2}$$-$$10^{-1}$ cm$^{-3}$ at 1 kpc to $10^{-4}$$-$$10^{-3}$ cm$^{-3}$ at 10 kpc. The number density profile of electrons in our simulations is described by the form:
 \be
 n_{e}=\frac{\rho}{\mu_{e}m_{H}}=\frac{n_{e0}}{r^{1.6}_{\text{kpc}}}, \label{ne}
 \ee where $\mu_{e}^{-1}$ is the average number of free electrons per nucleon, and $\mu_{e}\approx 1.17$ for solar composition,
 $m_{H}$ is the atomic mass unit, $n_{e0}$ is the electron number density at $1$ kpc, $r_{\rm kpc}=r$/1kpc.
 The value of $n_{e0}$ is $10^{-2}$ cm$^{-3}$ in the ``basic run'' (run A), and the density profile in 1$-$10 kpc is well within the observational range mentioned above, while beyond 10 kpc, the gas has little effect on {\it Fermi} Bubbles. More realistic forms of gravity and gas distribution were adopted in \citet{Guo1} and \citet{Guo2}. The simplified form used here would not influence the results significantly, since the difference of density distribution between the two forms is not so large in the bulge or halo. The difference becomes significant close to the Galactic plane but the {\it Fermi} Bubbles are far away from the Galactic disk.

The temperature of ISM is $9.2 \times 10^{5}$ K, which is determined by the velocity dispersion $\sigma$. The temperature is almost the same as that in \citet{Miller2013}.

One important massive structure exists in the Galactic center region, i.e., the Central Molecular Zone (CMZ, \citealt{Morris1996}). It is elongated along the Galactic plane, just surrounding Sgr A*, with total mass of several $10^{7}~\msun$. The length is 400 pc, and the height is 75 pc. As has been shown by \citet{Zubovas2012}, this structure has influence on the motion of winds from Sgr A*. It can collimate the winds to the perpendicular direction of the Galactic plane. In the simulation, the CMZ is set to be a torus-like structure located on the X-Y plane, with inner radius of 80 pc and outer radius of 240 pc. It is in hydrodynamic equilibrium, and the rotating velocity is $\sqrt{2} \sigma$. The ratio between the height and radius is set to be 0.15 in all runs. From our test simulations, we find that the ratio does not influence the results significantly when it increases from 0.15 to 0.25. The maximum thickness of CMZ is 72 pc, close to observational result. The density of CMZ is set to be a constant. The total mass of CMZ is set to be $2\times 10^{7}~\msun$. As mentioned in \citet{Zubovas2012}, CMZ can not be blown away by the winds because the ram pressure force impacted on CMZ is much smaller than the gravitational force. But the top and bottom parts of CMZ can be affected by Kelvin$-$Helmholtz (KH) instability and will form an interesting structure, which can explain X-ray observations (see \S4.3).

 The wind is injected from the inner boundary of the simulation, which has a height of 20 pc, and a width of 16 pc. The initial energy density of ISM around
 inner boundary is $\sim2\times 10^{-9}~{\rm erg~cm^{-3}}$. This pressure around the black hole supplies a threshold and only winds with ram pressure higher than this value will be able to push the ISM away and induce shocks in the
 galactic halo. In most runs of our model, the ram pressure of wind is about twice the initial pressure of ISM around the injection region.

\subsection{Equations}
The hydrodynamic equations describing the interaction process are as follows. Viscosity and thermal conductivity are included.
\begin{eqnarray}
\frac{d \rho}{d t} + \rho \nabla \cdot {\bf v} = 0,\label{hydro1}
\end{eqnarray}
\begin{eqnarray}
\rho \frac{d {\bf v}}{d t} = -\nabla P -\rho \nabla \Phi +\nabla
\cdot {\bf T},\label{hydro2}
\end{eqnarray}
\begin{eqnarray}
\frac{\partial e}{\partial t} +\nabla \cdot(e{\bf v})=-P\nabla \cdot
{\bf v}+{\bf T}:\nabla {\bf v}+\nabla \cdot(\kappa\nabla T), \label{hydro3}
   \end{eqnarray}
\begin{eqnarray}
{\bf T}=\mu (\nabla {\bf v}-\frac{2}{3} {\bf I} \nabla \cdot {\bf
v}).
\end{eqnarray}
 Here $\rho$ is the density of the gas, \emph{P} is the gas pressure, \emph{e} is the internal energy density of the gas, \textbf{v} is the velocity,
 {\bf T} is the viscous stress tensor, \textit{T} is temperature, $\mu$ is the viscosity coefficient, $\kappa$ is the heat conductivity coefficient, $\bf I$ is the unite tensor.
 The relationship between the gas pressure and the internal energy density is described by $P=(\gamma-1)e$. Radiative cooling is neglected. We have estimated the total energy lost by bremsstrahlung cooling within 10 Myr, and found that it is no more than a few percent of the total energy injected by wind.

\subsection{Viscosity}
The values of viscosity coefficient $\mu$ adopted in our models are shown in Table 1.  For comparison, we also run a model with $\mu=0$ (run C). As argued in \citet{Guo2}, the nature of viscosity is still highly uncertain.
  For a fully ionized, unmagnetized plasma, the dynamical viscosity coefficient is (\citealt{Spitzer62})
 \be
 \mu_{\rm visc}=6.0\times10^{3}\left(\frac{\text{ln }\Lambda}{37}\right)^{-1}\left(\frac{T}{10^{8}\text{ K}}\right)^{5/2} \text{  g cm}^{-1}\text{ s}^{-1}\text{,} \label{equvisc}
 \ee
where $\text{ln }\Lambda$ is the Coulomb logarithm. The viscosity coefficient is 2 $\text{g cm}^{-1}~\text{s}^{-1} $ for a typical temperature of $4\times 10^{6}$ K in the shocked ISM, and $2\times 10^{6} \text{ g cm}^{-1}~\text{s}^{-1}$ for $10^{8}$ K inside the bubble. In the present work, for simplicity we set the viscosity coefficient to be a constant which is very close to the value in the shocked ISM while quite different from that inside the bubble. But in the CMZ region we calculate the viscosity coefficient according to equation (\ref{equvisc}). CMZ gas will not be suffered from the effect of viscosity since the viscosity coefficient is very low there.

  \begin{figure*}[!htb]
    \centering
    \begin{center}
      \includegraphics[width=0.3\textwidth]{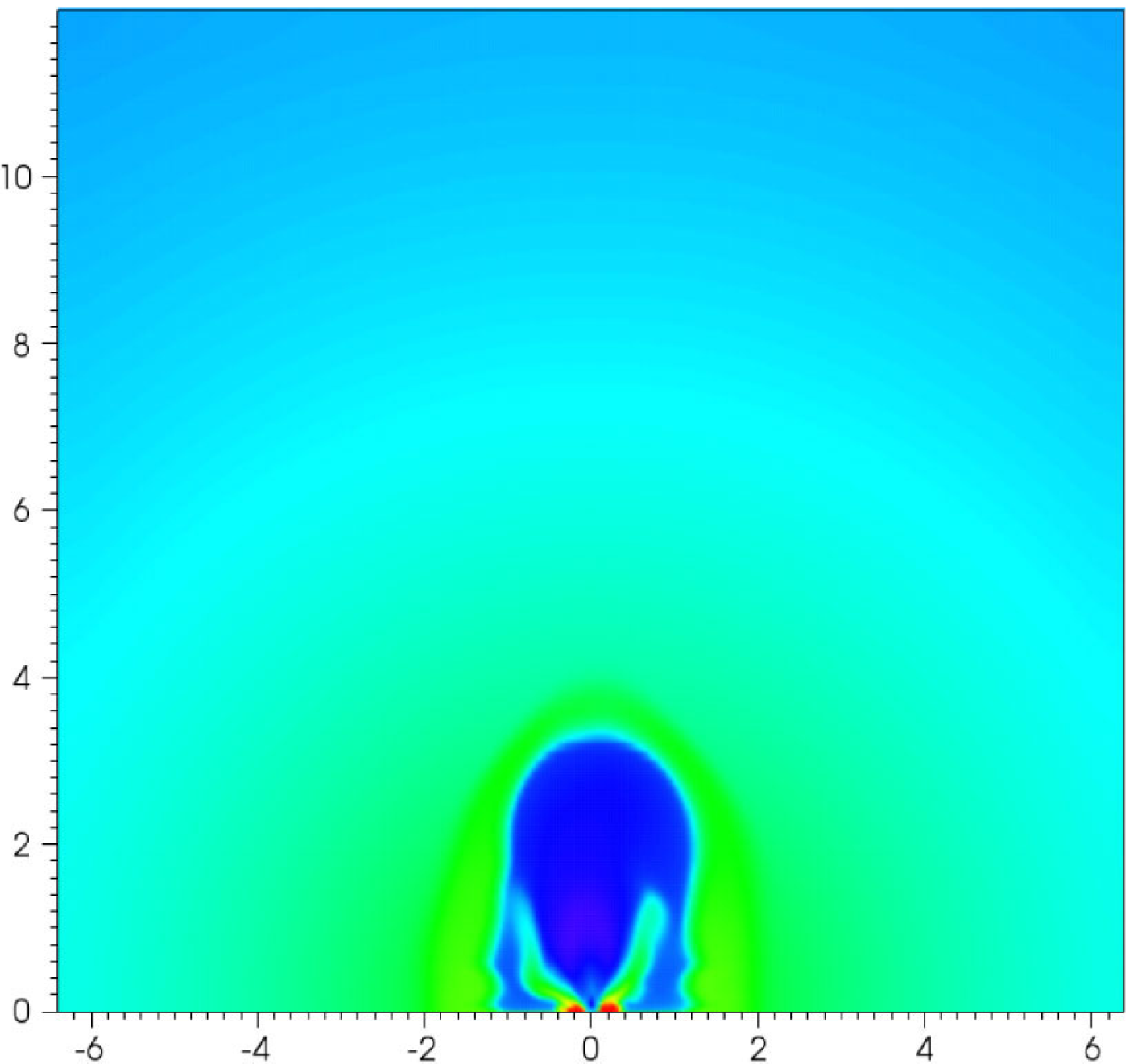}\hspace{0.27cm}
      \includegraphics[width=0.3\textwidth]{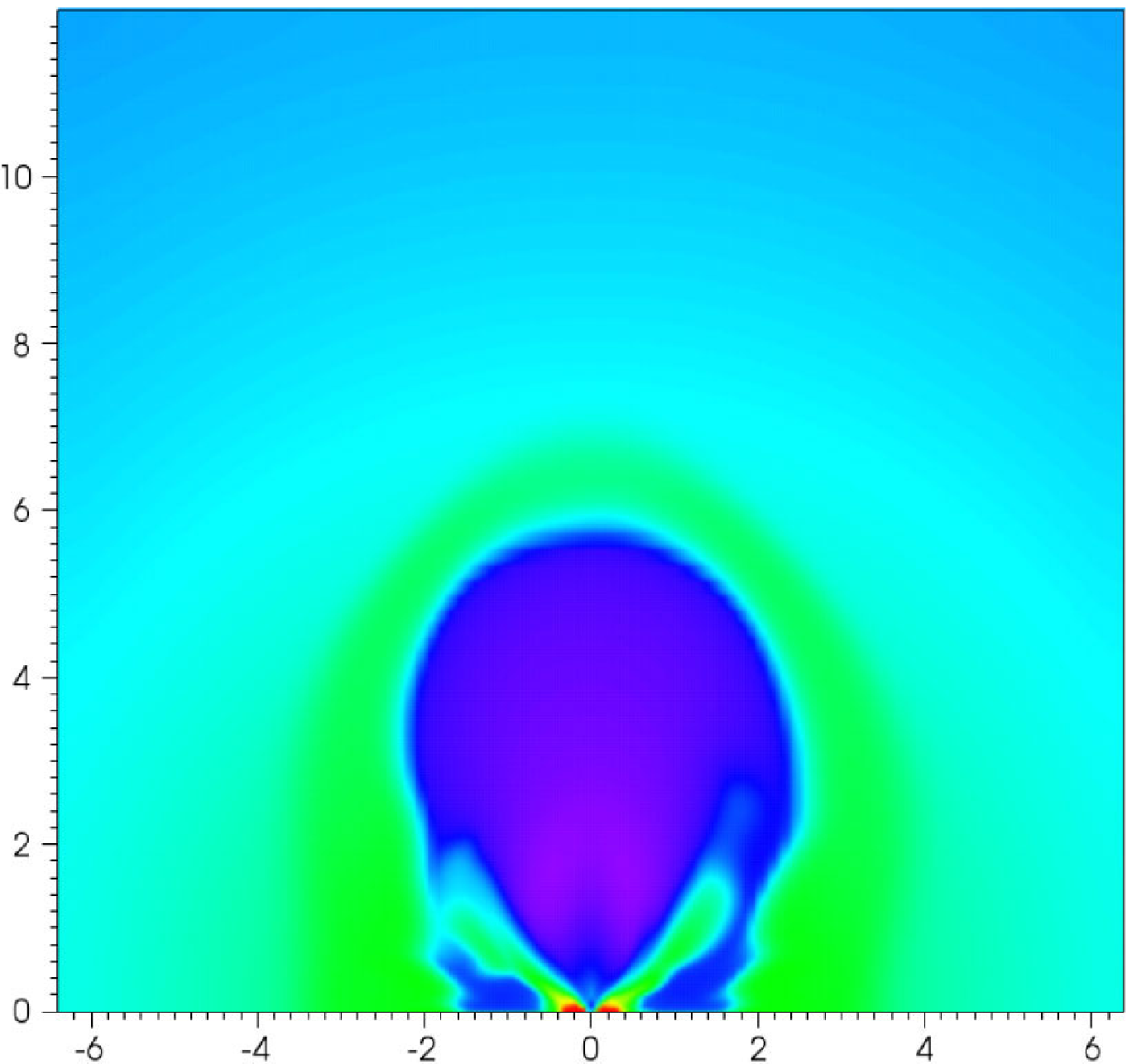}\hspace{0.27cm}
      \includegraphics[width=0.35\textwidth]{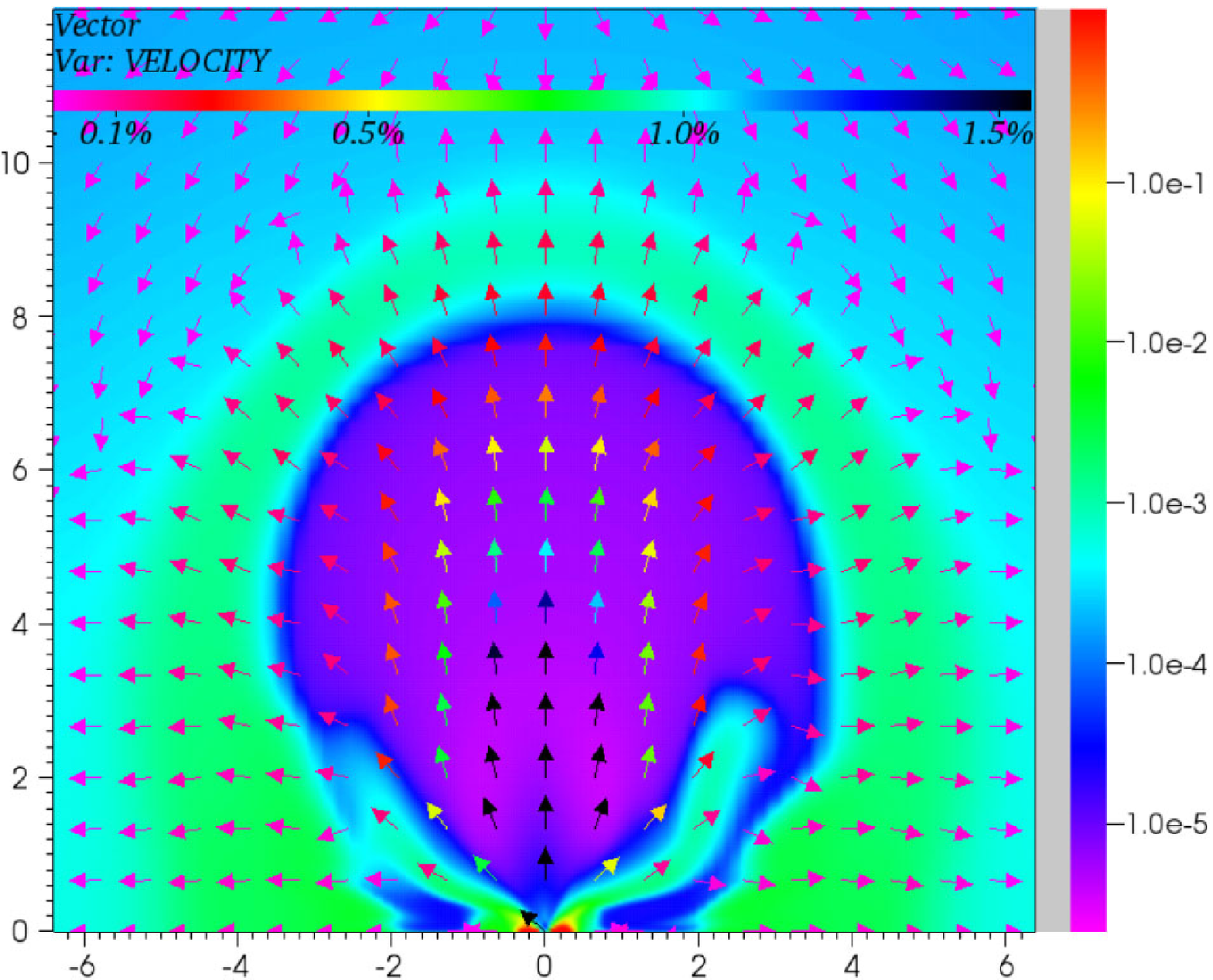}\hspace{0.27cm}\\
      \vspace{0.2cm}
    \end{center}
    \begin{center}
      \includegraphics[width=0.3\textwidth]{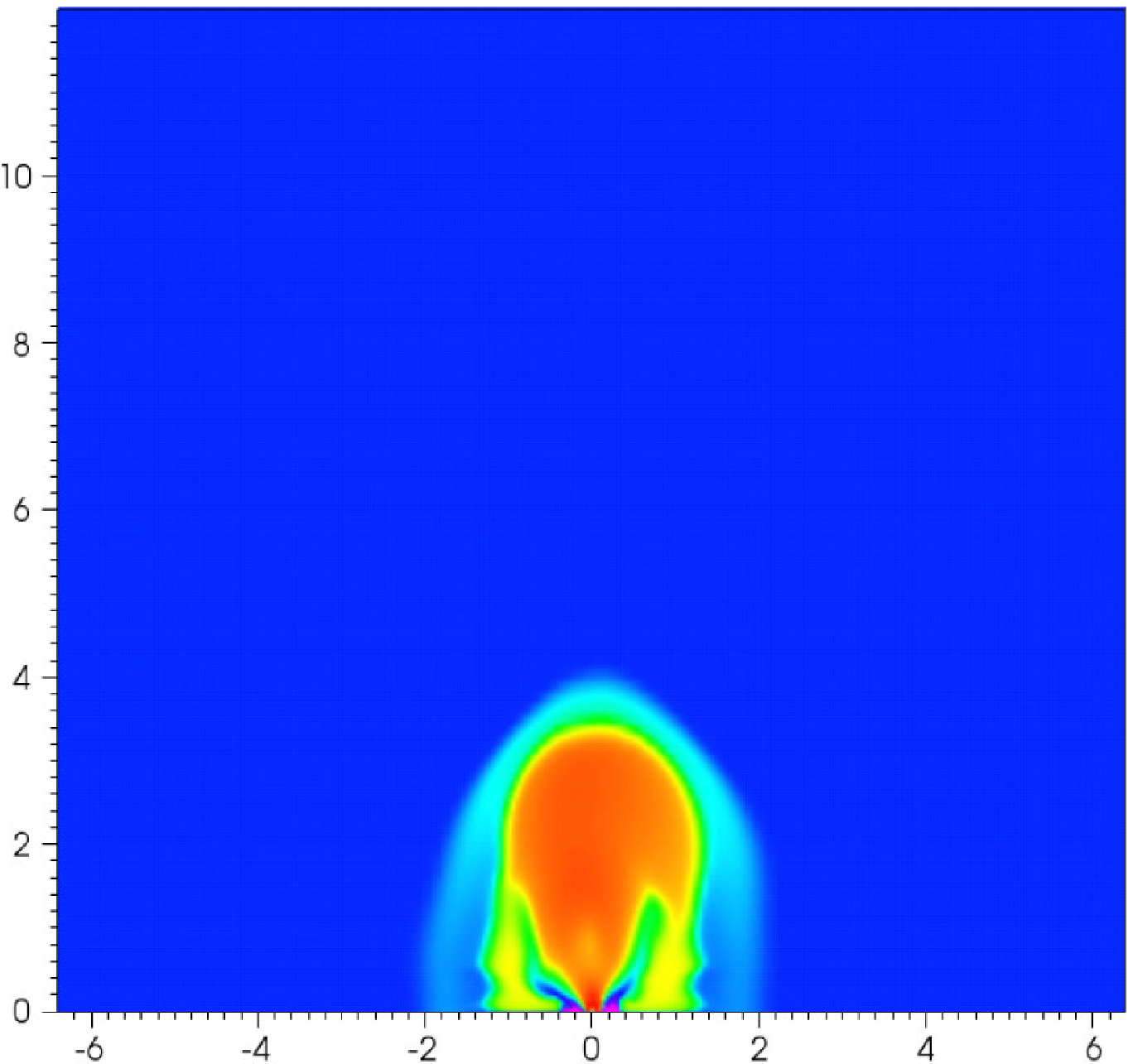}\hspace{0.29cm}
      \includegraphics[width=0.3\textwidth]{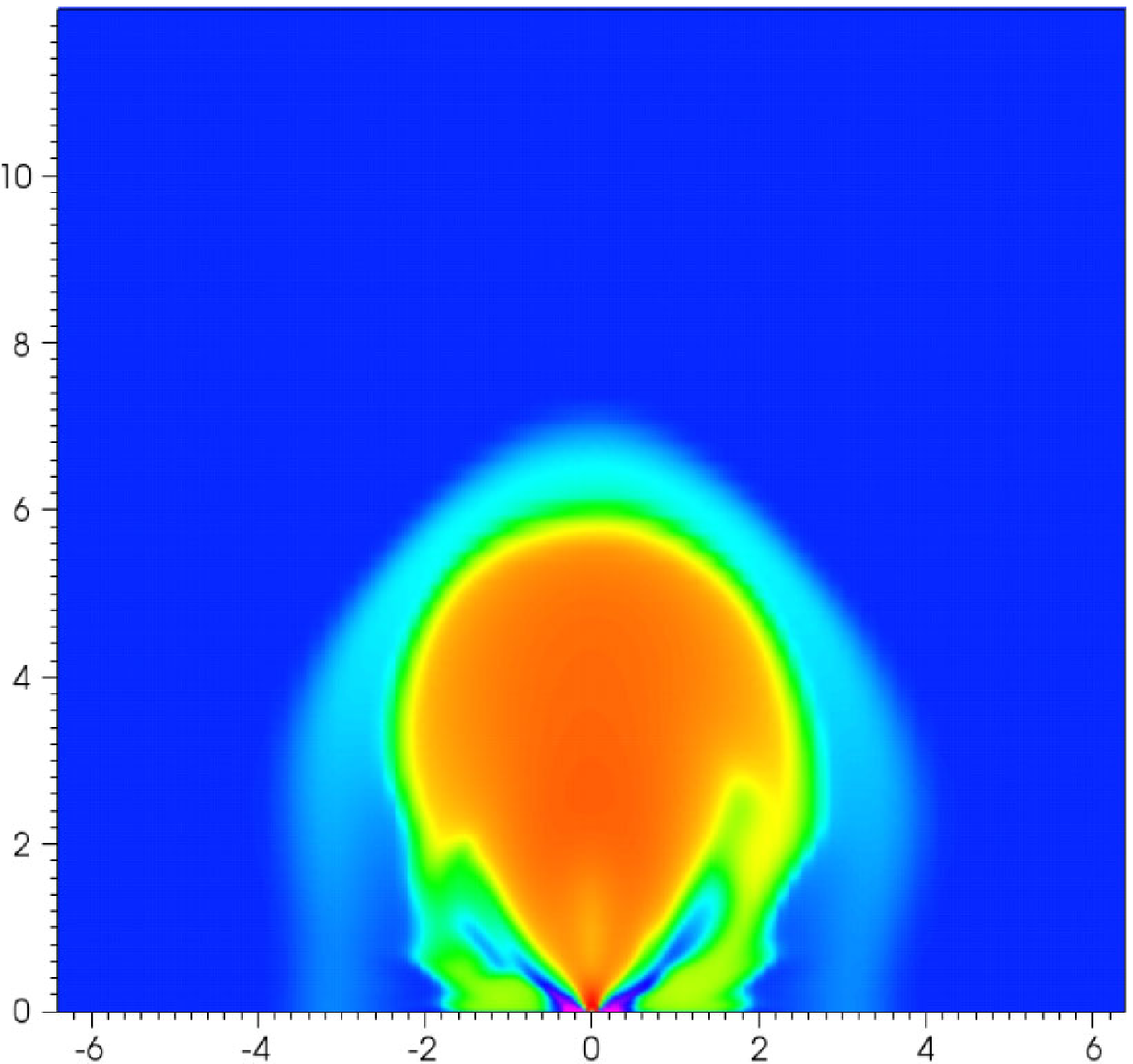}\hspace{0.29cm}
      \includegraphics[width=0.35\textwidth]{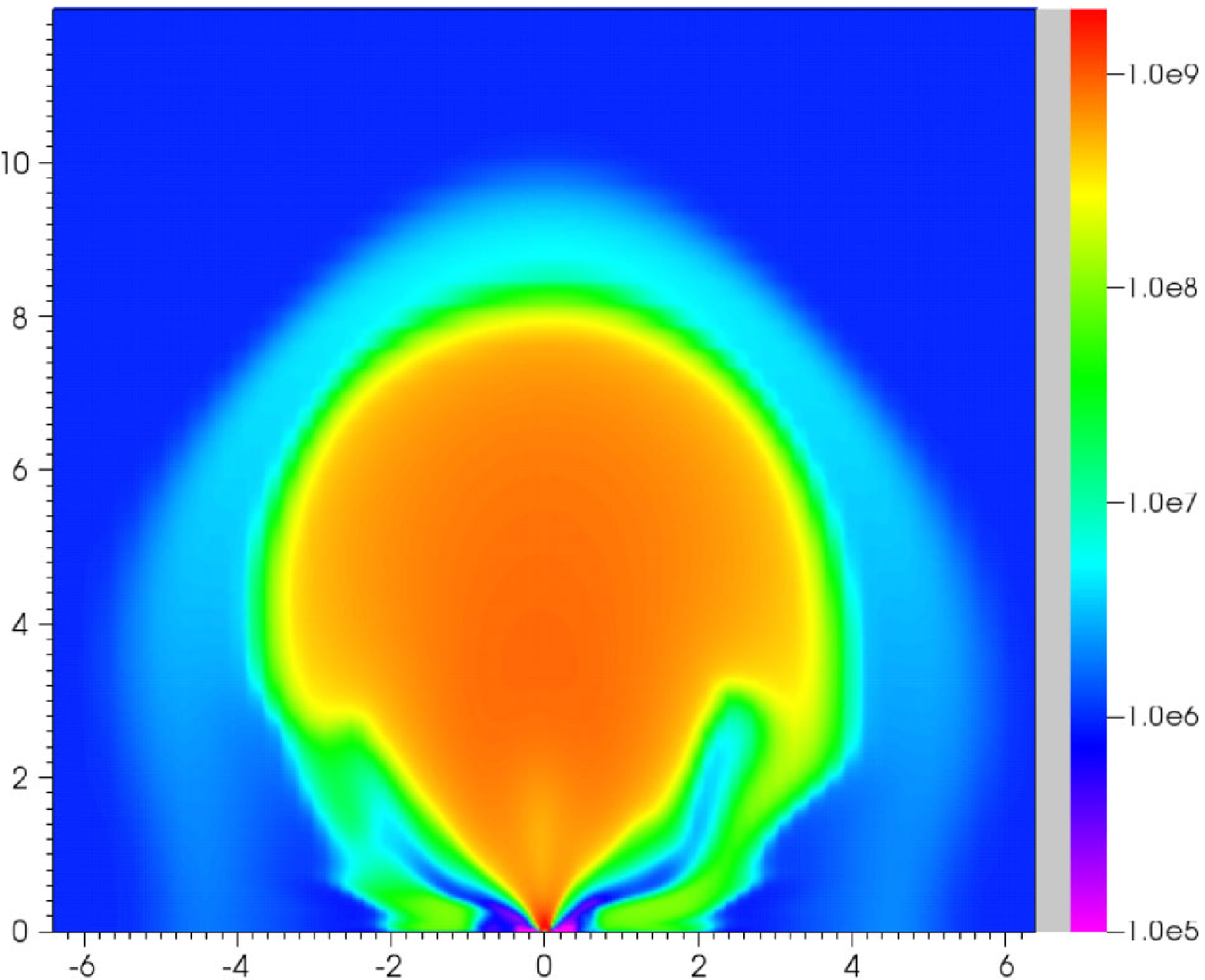}\hspace{0.29cm}
    \end{center}
    \caption{Evolution of morphology in run A (X-Z slice) for the number density (top panel) and temperature (bottom panel). From left to right, the plots correspond to $t=4$, 8, and $12.3$ Myr, respectively. The velocity field is added in the top right panel with values in unit of the light speed.}
   \label{label1}
 \end{figure*}

As pointed out by { \citet{Guo2}}, viscosity plays an important role because it can suppress instabilities so that we can get a smooth edge of the bubbles. The value of viscosity also influences the width of the bubble. We will discuss this point in more detail in \S4.4.

\subsection{Thermal Conductivity}
We also include thermal conduction in our models except run E. It makes the distribution of gas inside bubbles uniform. The heat flux $Q$ is given by:
  \be
   Q=-\kappa \nabla T,
  \ee here $\kappa$ is the coefficient of thermal conductivity. For a fully ionized gas, $\kappa$ is given by (\citealt{Spitzer62}):
  \be
   \kappa\approx 2\times 10^{-4} \frac{T^{5/2}}{Z^{4}\text{ln }\Lambda}  \text{  erg s}^{-1}~\rm{K}^{-1}~\rm{cm}^{-1}.\label{thercon}
  \ee
In reality, thermal conduction will be strongly affected by magnetic field. Specifically, in the direction perpendicular to the magnetic field,
thermal conduction will be strongly suppressed because it is difficult for the particles to move across the field lines. In addition, in a collisionless fluid,
  thermal conduction would be saturated, but the calculation of heat flux in this case is still on a phenomenological level with an artificially assumed factor (\citealt{Cowie1977}).
We find that our results are not sensitive to the value of $\kappa$. Even a value of the coefficient of thermal conductivity orders of magnitude lower than that determined by eq. (\ref{thercon}) is enough to smooth the distribution of gas within the bubble.

\section{RESULTS}
 \subsection{The morphology}

In our model, we identify the CD to be the edge of the observed {\it Fermi} bubbles. The region of shocked ISM  is the ``surrounding region''. When magnetic field is included, the field lines in this region will be aligned with the CD, which prevents the diffusion of relativistic electrons across the bubble edge. This then explains why the edge of the bubbles is so sharp (\citealt{Yang2012}). This mechanism also applies to our model since in reality magnetic field should exist.

\begin{figure*}[!htb]
     \includegraphics[width=0.41\textwidth]{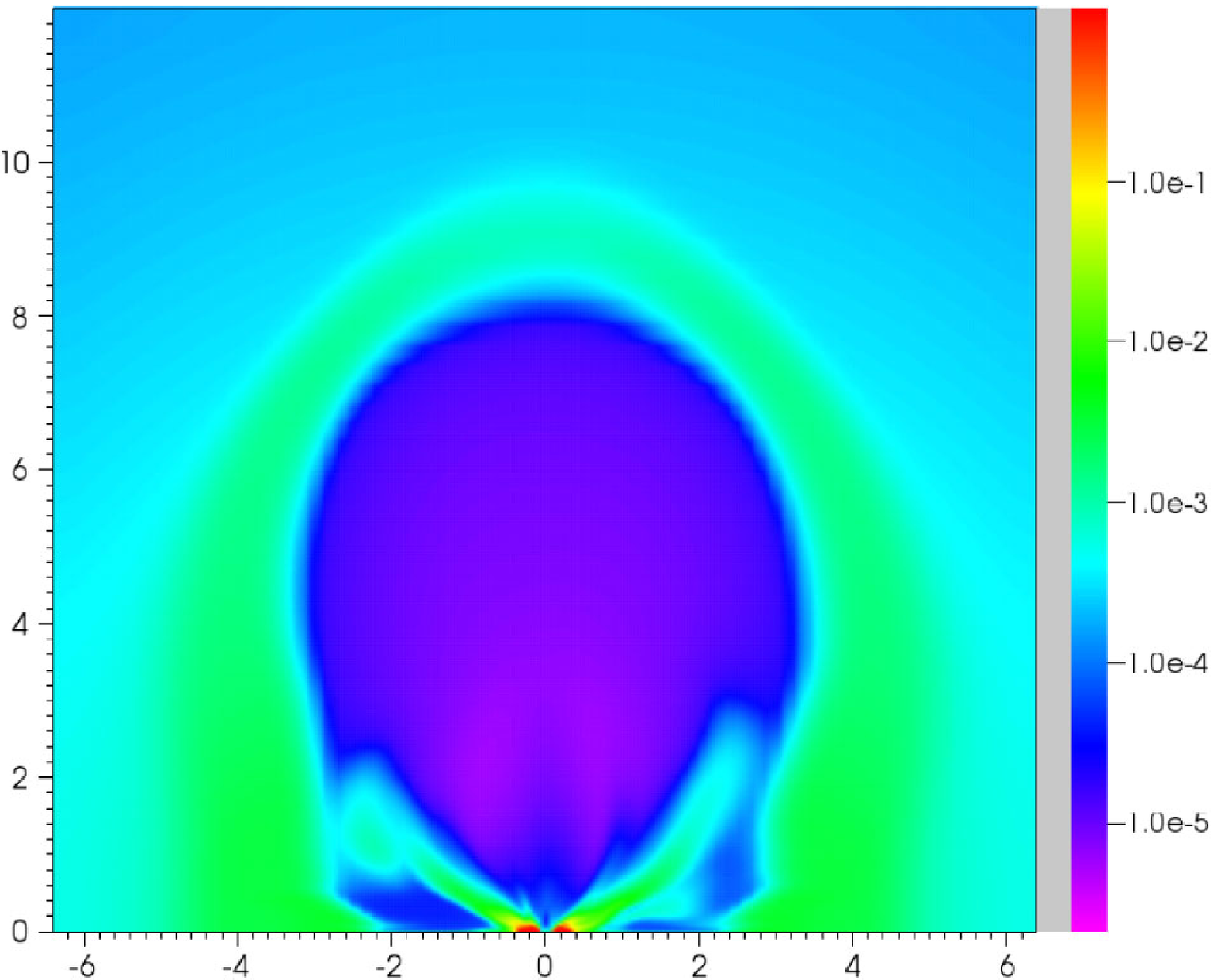}\hspace{2.0cm}
     \includegraphics[width=0.41\textwidth]{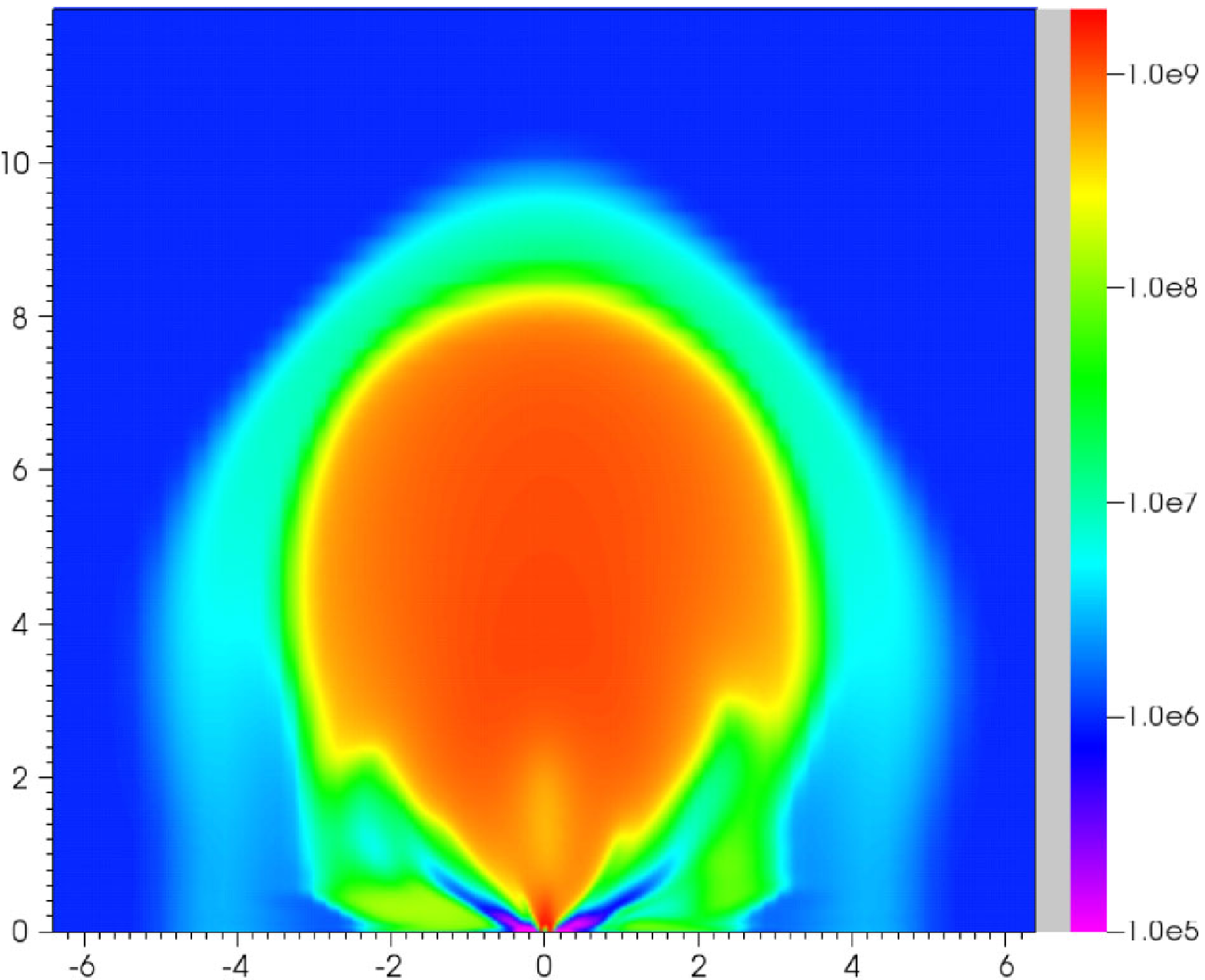}\hspace{1.2cm}
  \caption{The distributions of number density (left) and temperature (right) of electrons on $Y=0$ plane for run G (with a higher mass outflow rate than run A) at t=8.1 Myr. Compared with Fig. \ref{label1}, we can see that the density inside the bubble is larger, and the temperature in the whole region is higher.}
  \label{label8}
  \end{figure*}

If we only want to explain the morphology of the bubbles, we find that we have relatively large freedom in terms of the values of velocity and mass flux of winds. For example, we can use a smaller wind velocity and a higher mass outflow rate (see model G, or Fig. \ref{label8}), or a larger wind velocity and a lower mass outflow rate to get the ``correct'' morphology. However, observations put additional constraints, such as temperature. We choose run A as our ``basic run'' because not only the morphology but also other properties of the bubbles are consistent with observations. In run A, winds need to last for 12.3 Myr to get the ``correct'' morphology of the bubbles, as shown by Fig. \ref{label1}. The height and width of the bubble are 8 kpc and 7 kpc respectively,
which corresponding to a projected bubble with a latitude of $50^{\circ}$ and a longitude of $50^{\circ}$. Although the wind is set to be injected isotropically, the massive CMZ surrounding Sgr A* blocks the lateral movement of winds, and forces them to move
upwards. In other words, the CMZ successfully collimates the wind to the perpendicular direction of the Galactic plane. This is why we can obtain a bubble with narrow waist near the Galactic plane, instead of a hemispherical bubble buckling on the Galactic plane.

\begin{figure*}
   \centering
    \begin{center}
      \hspace{0.25cm}\includegraphics[width=0.295\textwidth]{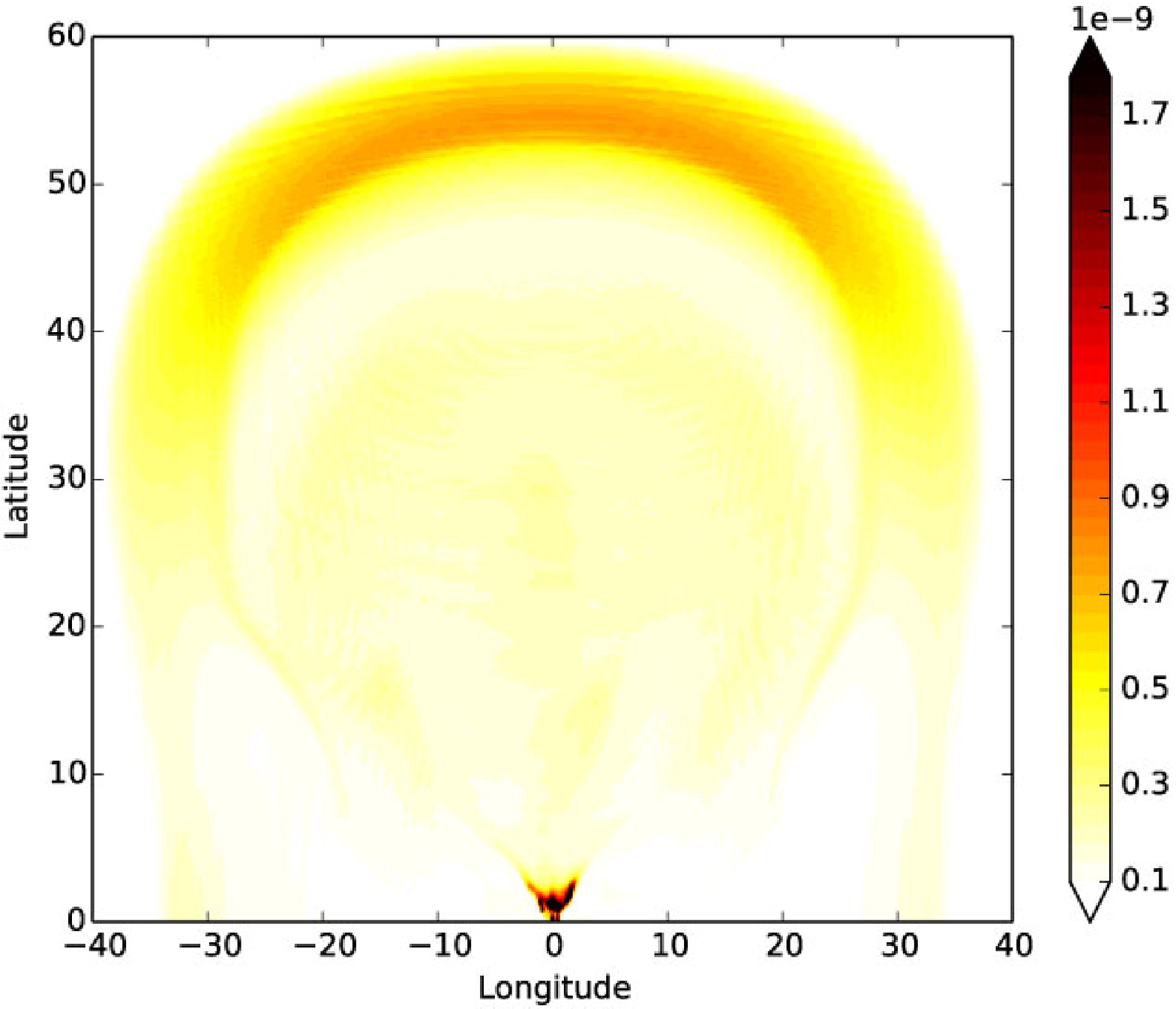}\hspace{0.33cm}
      \includegraphics[width=0.295\textwidth]{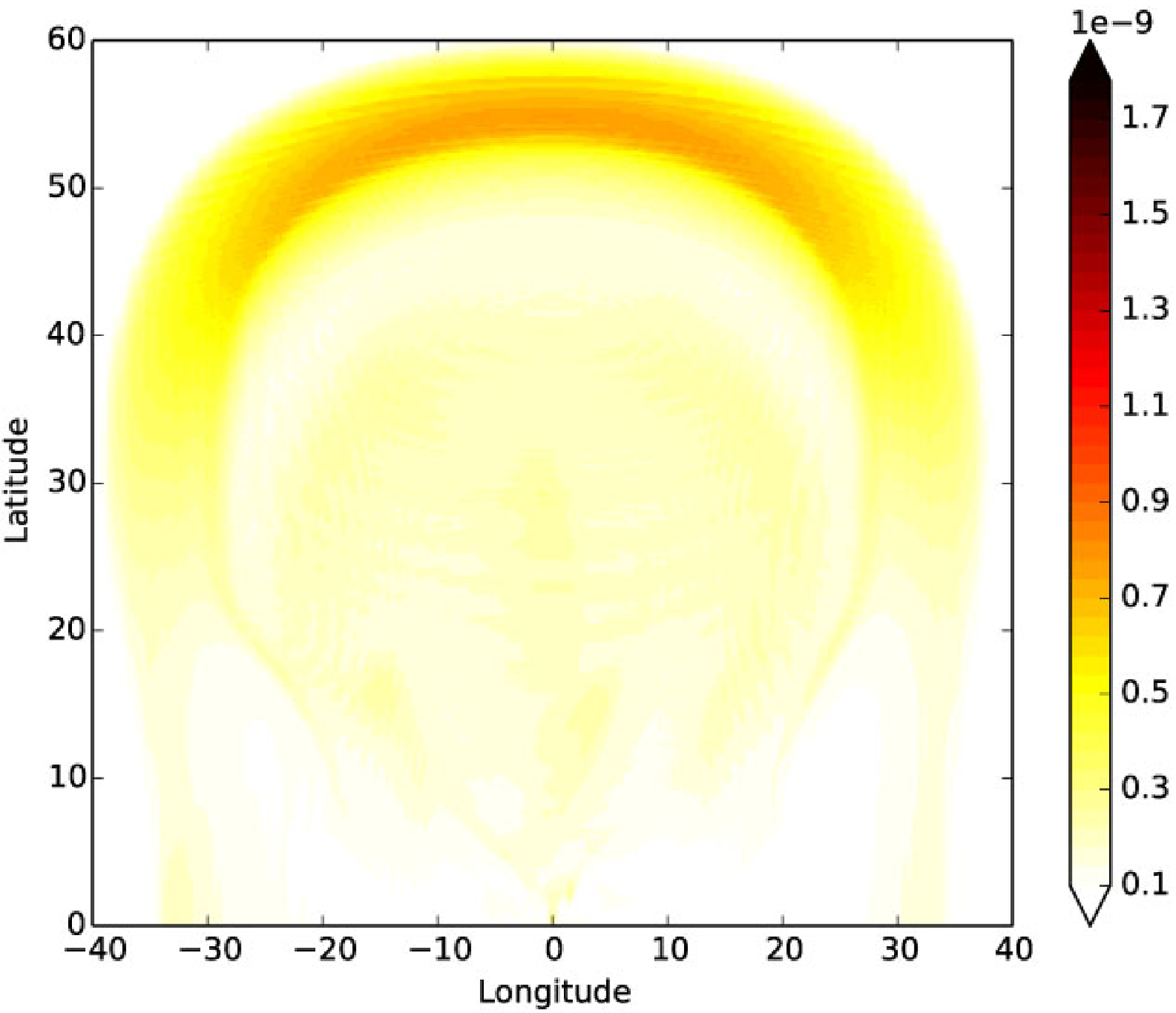}\hspace{0.3cm}
      \includegraphics[width=0.295\textwidth]{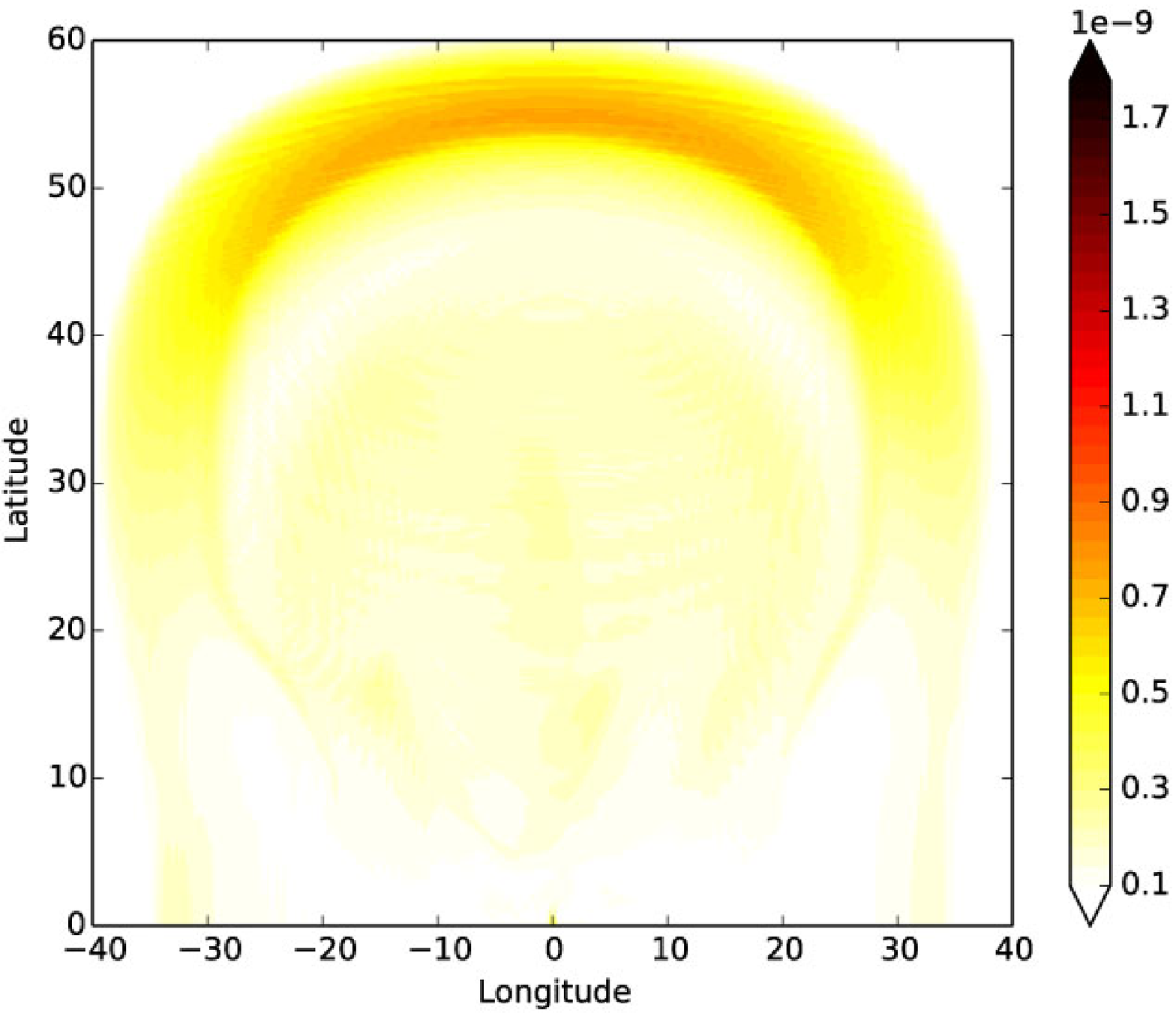}
    \end{center}
    \begin{center}
      \includegraphics[width=0.32\textwidth]{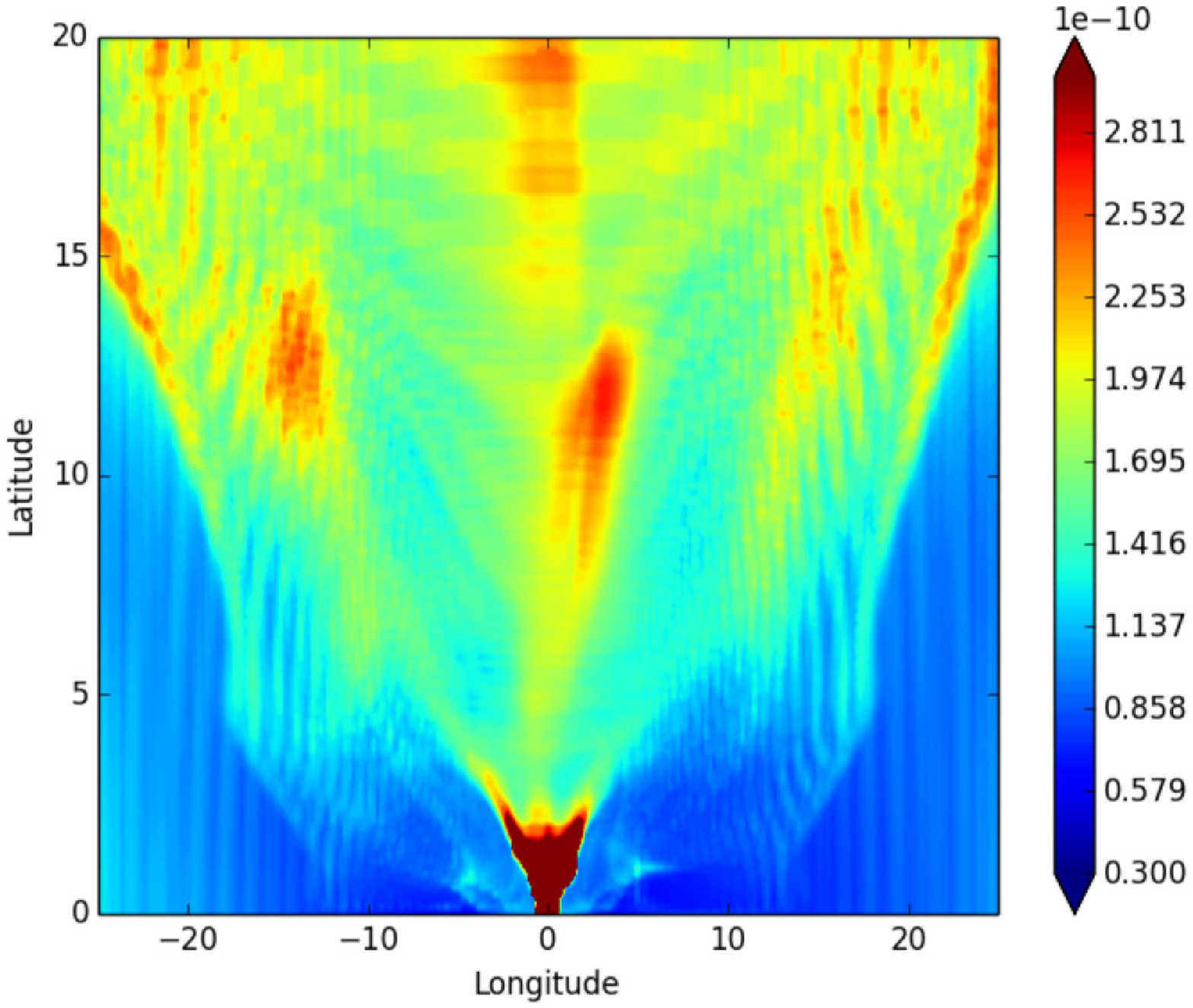}
      \includegraphics[width=0.32\textwidth]{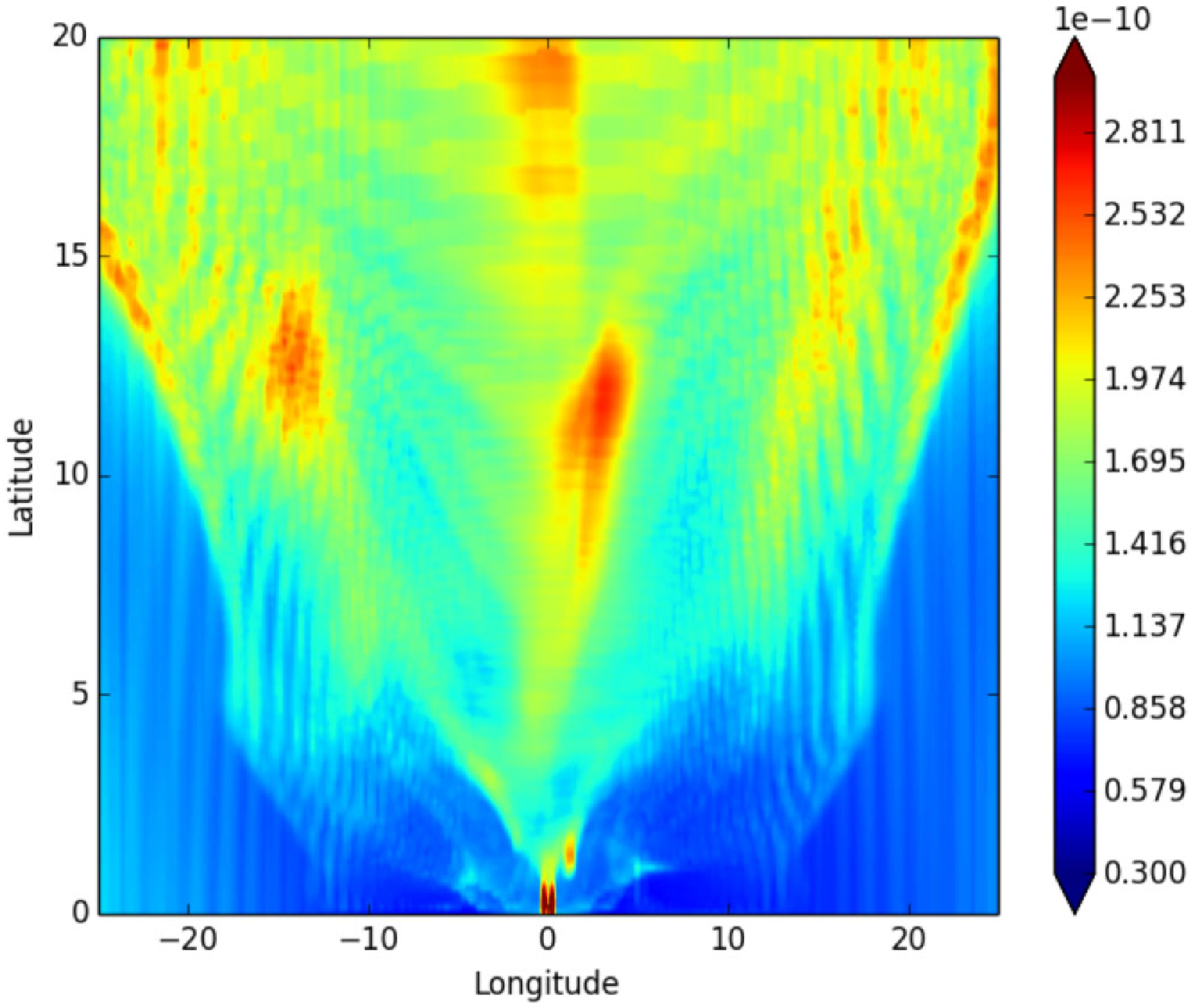}
      \includegraphics[width=0.32\textwidth]{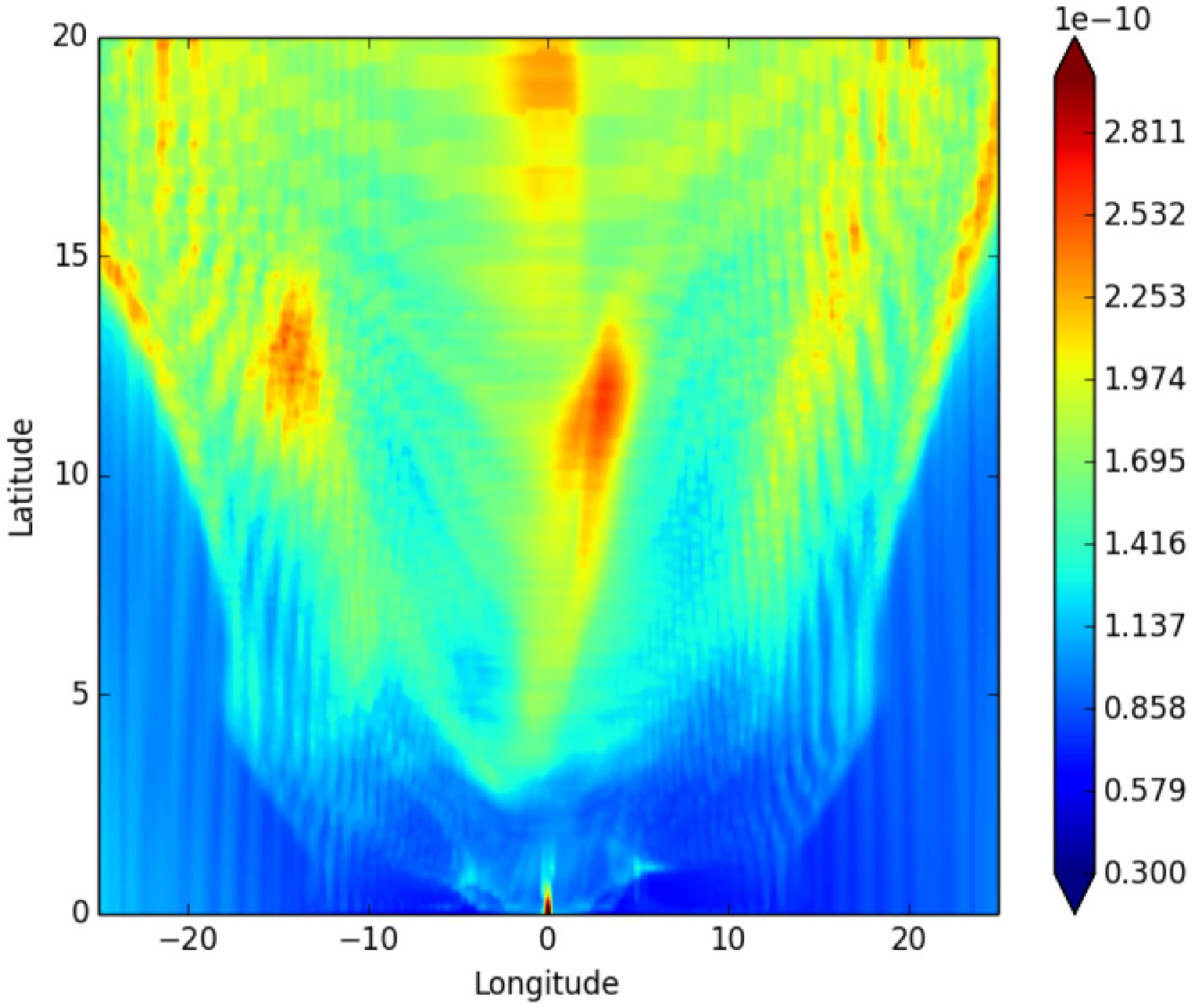}
    \end{center}
   \caption {The X-ray structure in R6+R7 band (0.5 keV$-$1.5 keV) obtained from run B.
   Top and bottom panels are for different spacial scales, with the bottom one zooming in the center part near the Galactic Center (GC). For each panel, from left to right, the plots correspond to different time durations ($\delta t$) from the quenching of the Sgr A* activity, with $\delta t=0.1$, 0.2 and 0.3 Myr, respectively. The brightness of the inner conical structure gradually dims out. For the bottom-left plot, both the limb-brightened surrounding structure outside bubble and the conical structure near the GC are clearly seen, in good consistency with observations (\citealt{Snowden1997}; \citealt{Wang2002}).  For the middle one, this structure is significantly weaker; while for the right one, the structure in $|b|\la 5^{\circ}$ begins to disappear. }
  \label{label3}
 \end{figure*}

We have tried to explore when the activity of Sgr A* quenched and entered into the quiescent state by running ``run B''. We find that only if the quiescent time is shorter than 0.2 Myr, the result will not be affected, i.e.,  showing a significant conical structure in X-ray band with latitude $|b|\la 10^{\circ}$ (see lower panels in Fig. \ref{label3}). Because the quiescent timescale is so short compared with the age of {\it Fermi} Bubbles, although in all other runs in this work Sgr A* do not enter into the quiescent state as it should be, our simulation results will not be affected.

\subsection{Energy, Mass and Temperature}

For run A, the total energy injected by Sgr A* is about $7.7~\times 10^{55}$ erg. This energy is comparable to the injected energy from Galactic center estimated from some observations to some structures, such as the NPS structure, GCL, and EMR (\citealt{Totani2006}). The total internal energy of {\it Fermi} bubbles in our simulation is $ 2.2\times 10^{55}$ erg, which is consistent with the  observational value. The total kinetic energy of the {\it Fermi} Bubbles is only $2\times 10^{54}$ erg, much smaller than the internal energy. This is because the speed of the gas inside the bubble is subsonic.

 \begin{figure}[!htb]
   \centering
   \plotone{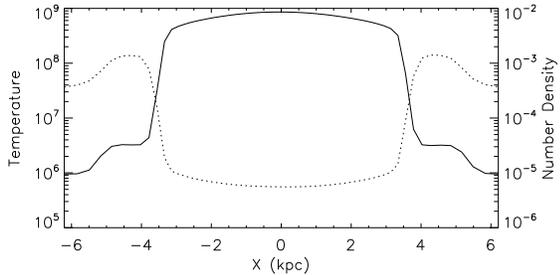}
   \caption{Profiles of temperature (solid line) and electron number density (dotted line) along the X-axis for run A at $t=12.3$ Myr, averaged from $z=4.5$ kpc to 6.0 kpc. Temperature is in unit of Kelvin, while number density in unit of ${\rm cm^{-3}}$. From this figure, we can see that the average temperature inside the
    bubble is $\sim 6 \times 10^{8}$ K, while in the surrounding region it is $\sim 3 \times 10^{6}$ K.}
   \label{plot2}
  \end{figure}

The total mass inside the bubble is about a few times $10^{5}~\msun$, which is much lower than the estimation of $10^{8}~\msun$ based on the assumed upper-limit of an average density $n\sim 10^{-2}$ cm$^{-3}$ in \citet{Su10}. This is because the density in our simulation is about three orders of magnitudes lower than the assumed value in \citet{Su10}, as shown by Fig. \ref{plot2}. We have the following comments to this ``discrepancy''. Firstly, our observational constraint on the density is poor, thus the total mass of $10^{8}~\msun$ is subject to a large uncertainty. Secondly, since the coefficient of thermal conductivity adopted in our work is low, we may have underestimated the evaporation process which may play an important role in transporting mass from the surrounding gas, including CMZ and shocked ISM, into the Bubbles. Thirdly, we assume the ISM homogeneous for simplicity, while in reality the ISM is likely to be clumpy. The dense clouds may be difficult to be blown away by winds so they will stay inside the bubble, which will significantly increase the mass of the gas within the bubbles.

  \begin{figure*}[!htb]
    \centering
    \begin{center} \hspace{1.0cm}
      \hspace{-1.0cm}\includegraphics[width=0.41\textwidth]{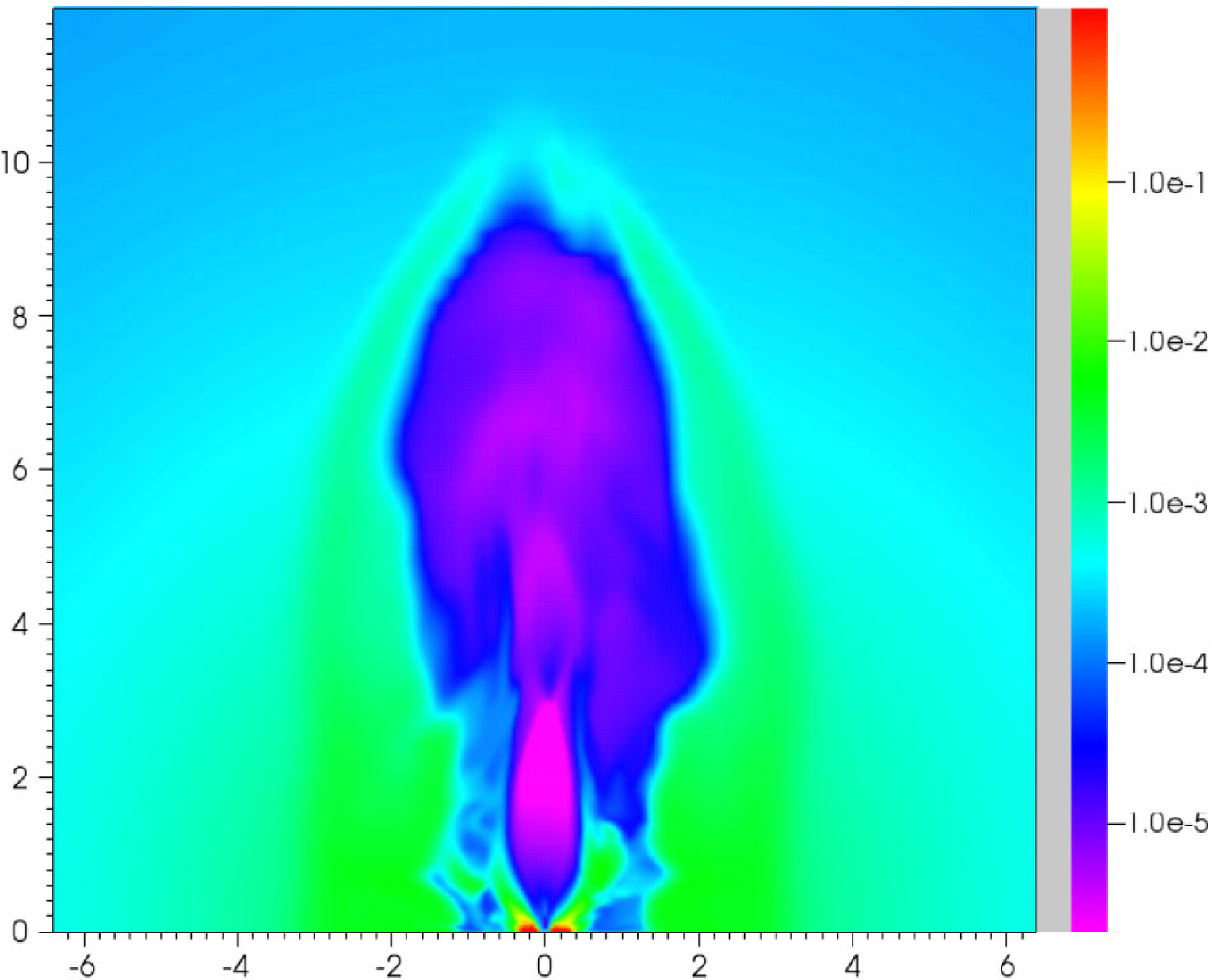}\hspace{3.cm}
      \includegraphics[width=0.41\textwidth]{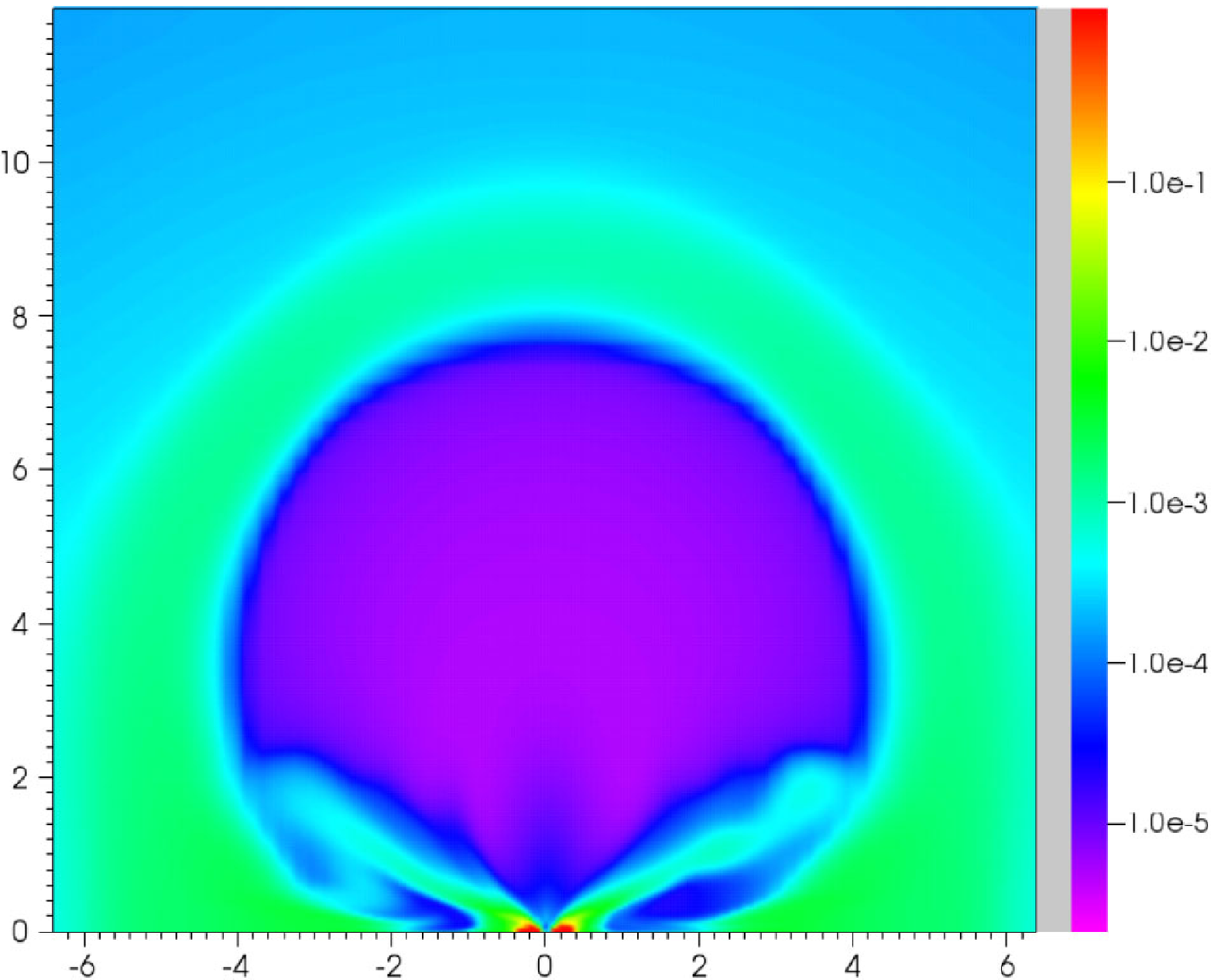}
      \vspace{0.3cm}
    \end{center}
    \caption{The effect of viscosity coefficient on the morphology of bubbles. The left and right plots show the number density of electrons for run C (without viscosity) at $t=7.6$ Myr and run D (with large viscosity) at $t=14.5$ Myr, respectively (X$-$Z slice). When the viscosity coefficient is higher, the bubble becomes
    more spherical.}
   \label{label4}
 \end{figure*}

The temperature of the gas is determined by the following equation:
  \be
  T=\frac{(\gamma-1)e\mu m_{H}}{k \rho},
  \ee where $T$ is temperature, $\mu$ is the molecular weight, which is 0.61 for solar composition. Temperatures of different runs are given in Table 2. In general, the temperature inside the bubble is several times $10^{8}$ K. Although there are some new results of Milky Way's hot
  halo recently, the temperature inside the {\it Fermi} Bubbles is still lack of data. So our result can be regarded as a prediction. We will discuss the temperature in the surrounding region between the CD and the forward shock in \S4.3.

\subsection{X-ray Structure}

 \begin{figure*}[!htb][here]
  \hspace{-0.8cm}\includegraphics[width=0.41\textwidth]{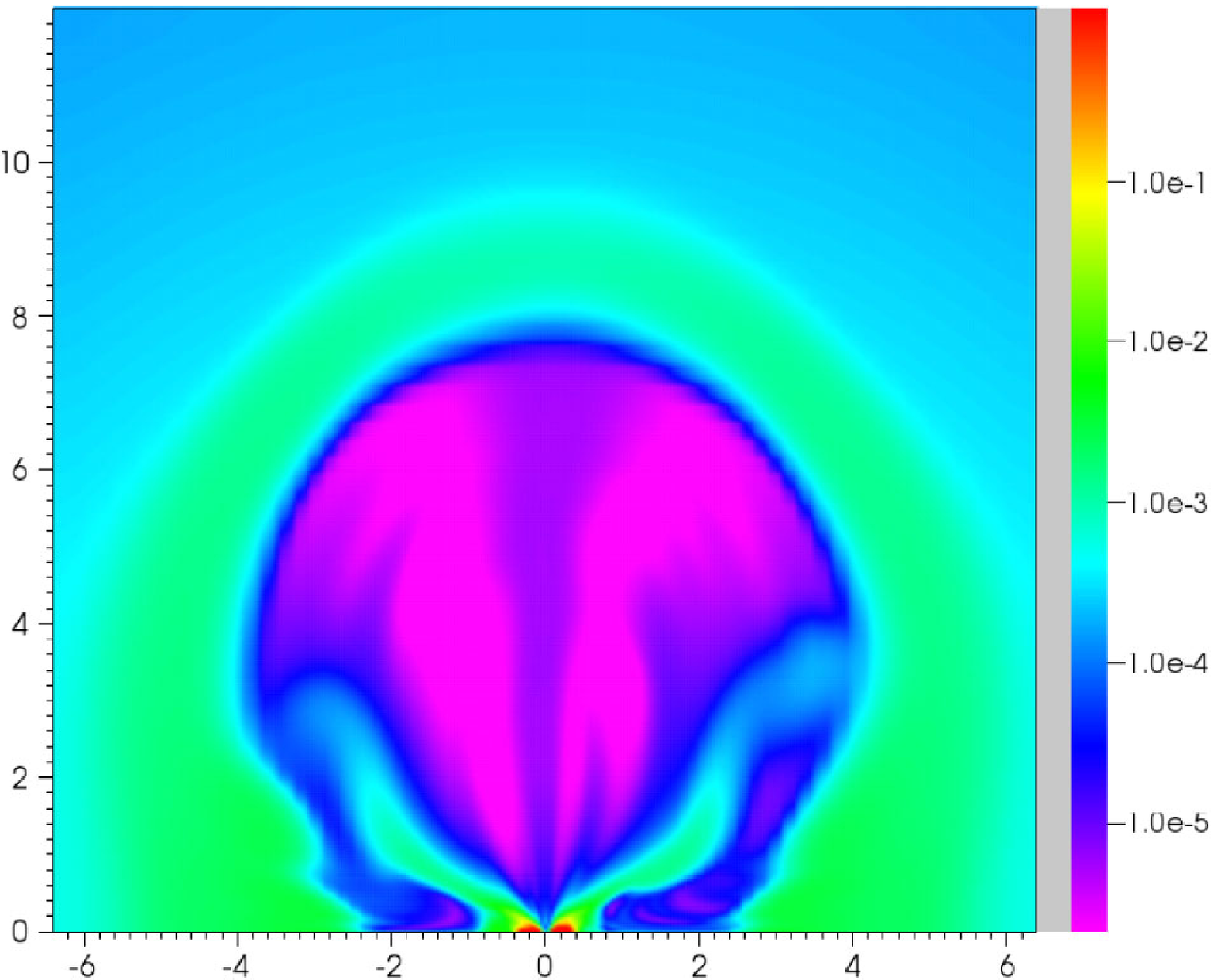}\hspace{3.1cm}
  \includegraphics[width=0.41\textwidth]{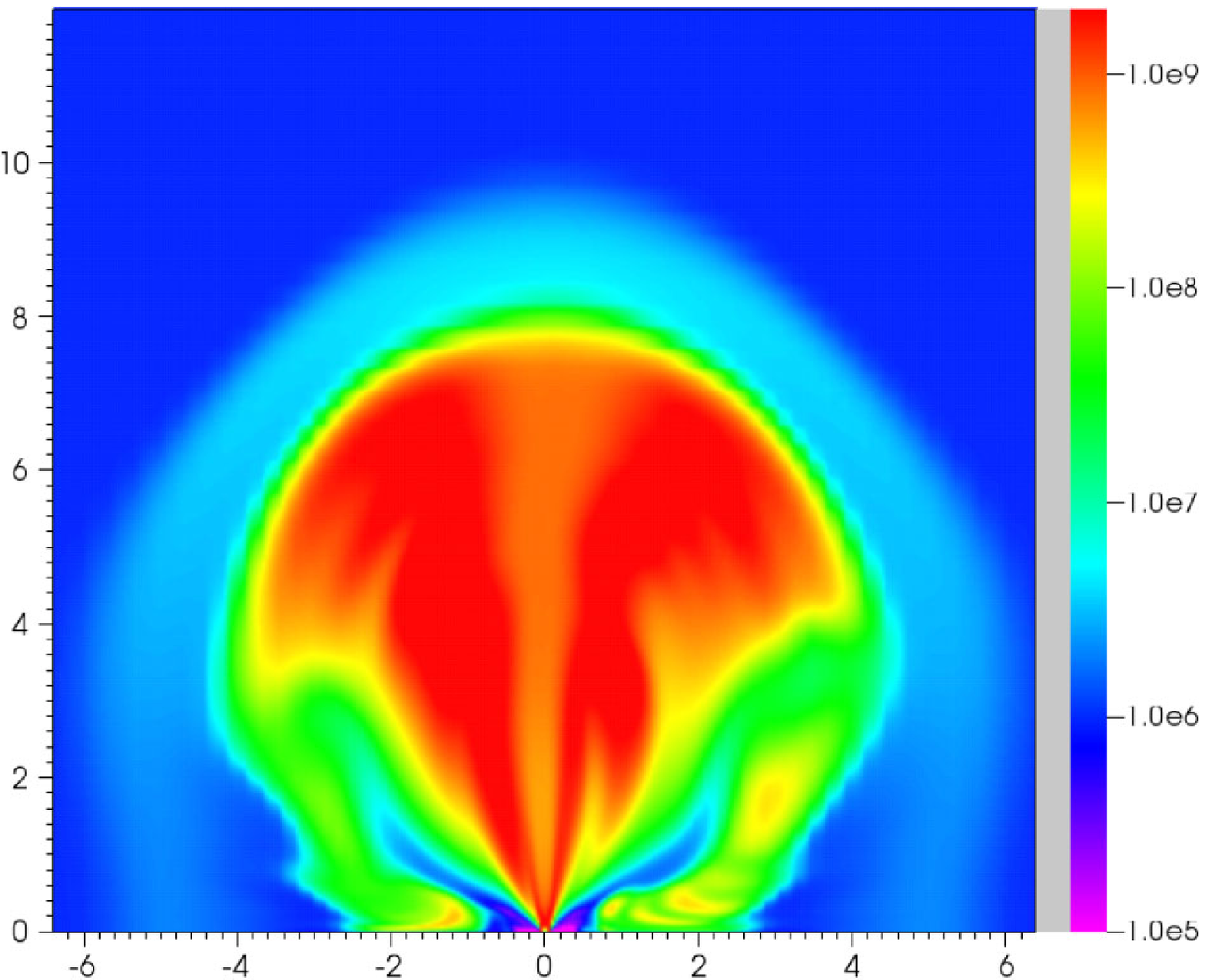}
  \caption{The distributions of electron number density (left) and temperature (right) at the $Y=0$ plane for run E (without thermal conduction) at $t=13.4$ Myr. Compared with Fig. 1, We can see that there is a ``jet-like'' feature with lower temperature and higher density through the middle of the bubble. This feature disappears when thermal conduction is included. }
  \label{label6}
 \end{figure*}

 \begin{figure*}[!htb]
  \includegraphics[width=0.41\textwidth]{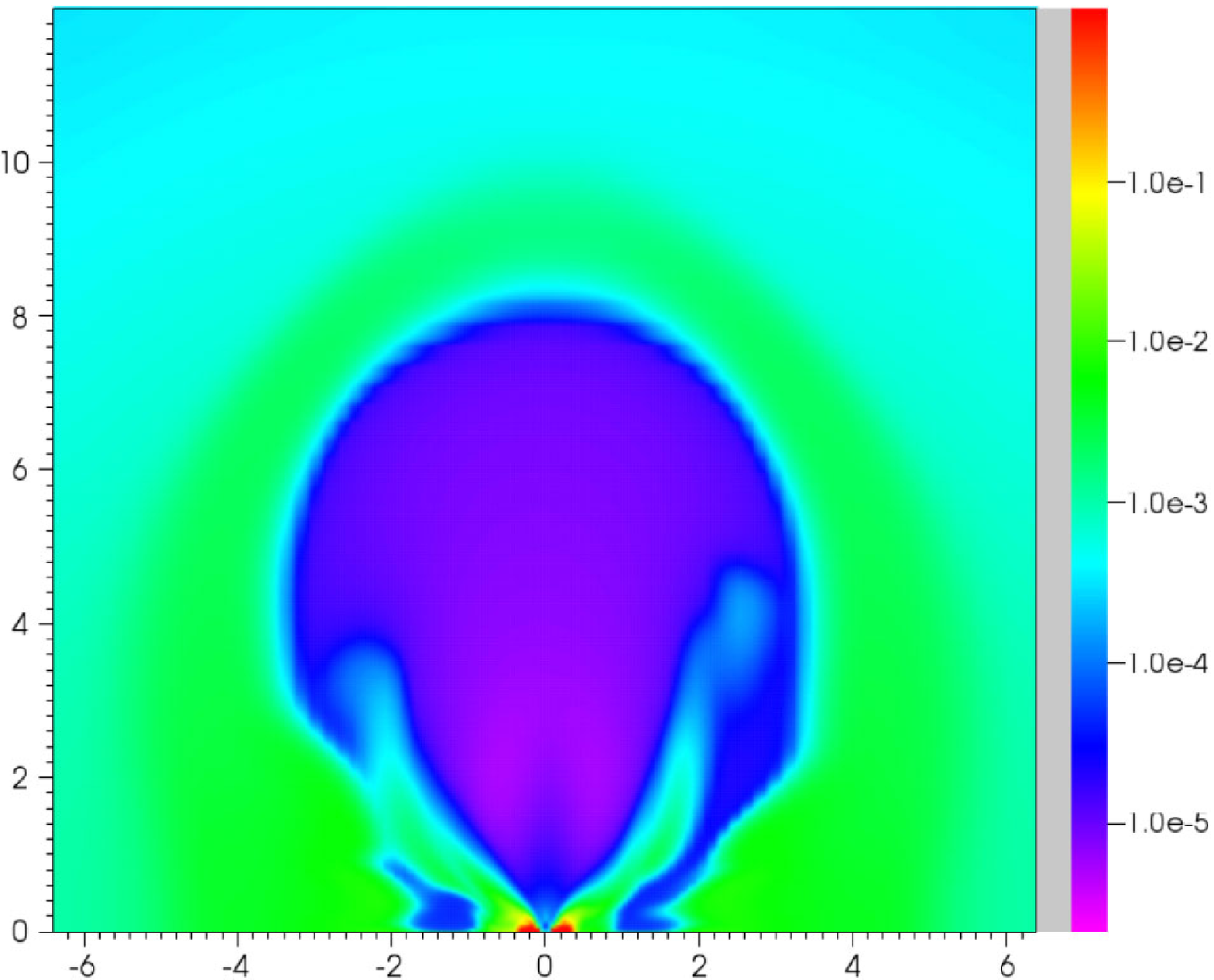}\hspace{2.0cm}
  \includegraphics[width=0.41\textwidth]{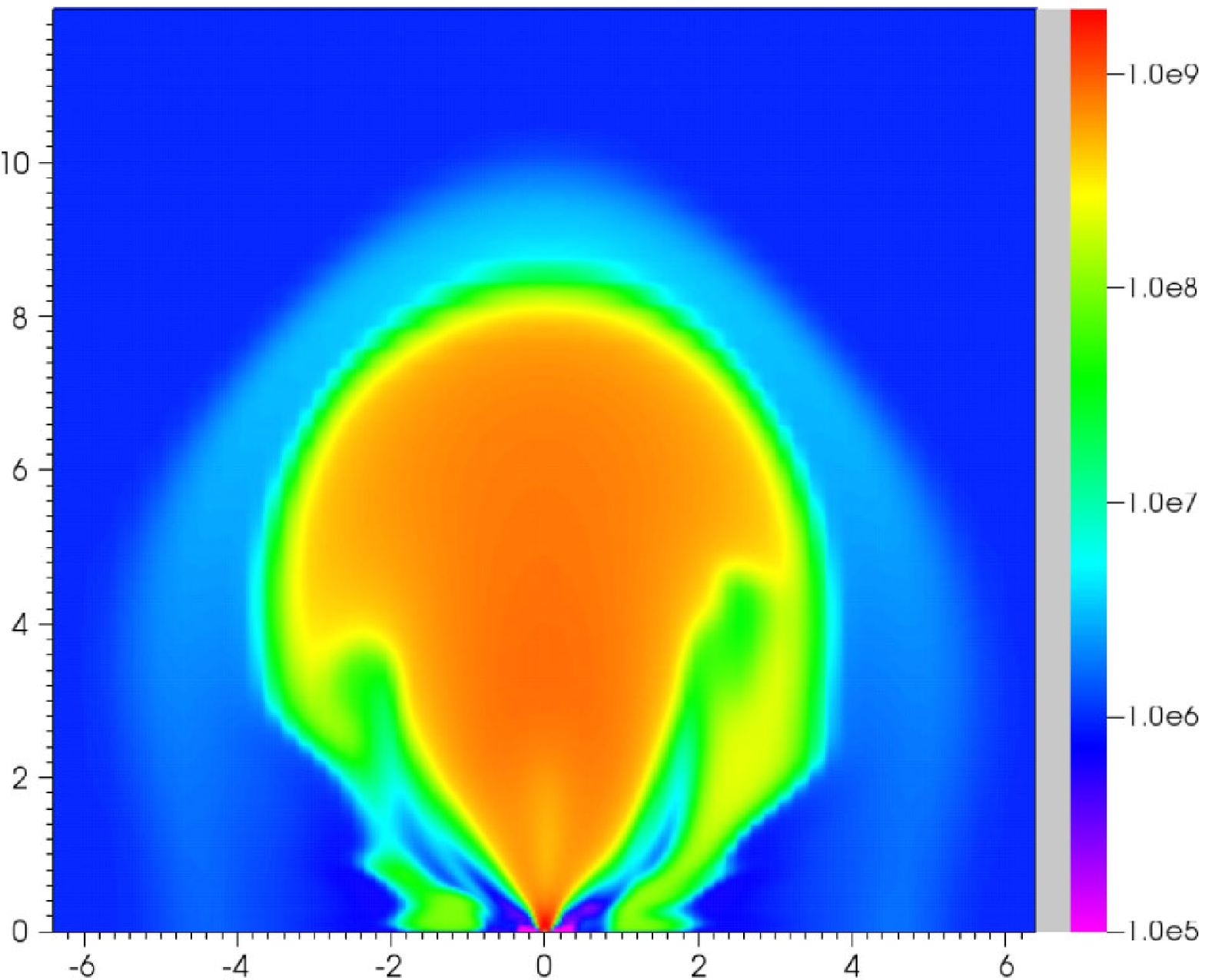}\hspace{1.2cm}
  \caption{The distributions of density (left) and temperature (right) on $Y=0$ plane for run F at $t=14.9$ Myr. In this model, the density of the ISM is two times higher than run A. In this case, more time is needed to form the bubbles. Both the density and the temperature inside the bubble are higher while the temperature of the surrounding structure is lower because of the lower speed of the forward shock.}
  \label{label7}
 \end{figure*}

We have calculated the predicted X-ray image by considering the bremsstrahlung radiation. Fig. \ref{label3} shows the result. We can see that the morphology is consistent with the limb-brightened X-ray structure obtained in {\it ROSAT} observations (\citealt{Snowden1997}; \citealt{Su10}). This structure corresponds to the shocked ISM. Here, we only calculate the bremsstrahlung radiation, so our images are only for qualitative comparison. The temperature inside the bubble is two orders of magnitude higher than the surrounding region while the density is two orders of magnitude lower. Therefore the outer region is much brighter than the interior of the bubbles in X-ray band.

Recent {\it Suzaku} observations have revealed that the temperature of the surrounding region in high latitude ($\ga +40^{\circ}$) is around 0.3 keV and the thermal pressure is $2\times 10^{-12}$ dyn cm$^{-2}$ (\citealt{Kataoka2013}). Our model is in good consistency with their results.  In run A, the temperature and thermal pressure in the same location are 0.4 keV and $1.2~\times~10^{-12}$ dyn cm$^{-2}$ respectively (also see Table 2, note that the temperature is space-averaged value of the shocked ISM). In contrast, in jet model and quasar outflow model, the predicted temperature is larger than a few keV (\citealt{Guo1}; \citealt{Yang2012}; \citealt{Barkov2014}) and 1 keV (\citealt{Zubovas2012}) respectively. We think that the main reason for such a discrepancy is that the wind velocities, and more importantly the mass fluxes of wind, in these two models are too high.

We have calculated the bremsstrahlung radiation, and found that the total lost energy is no more than ten percent of the total internal energy of the surrounding region. So cooling effect is very weak. We note that the {interaction region} between the winds and the CMZ gas is also quite bright in X-ray band, which looks like a cone upside down on the Galactic plane. This structure explains the features observed in the {\it ROSAT} X-ray survey by \citet{Snowden1997} and \citet{Wang2002}.

As we have mentioned in \S4.1, the brightness of the conical X-ray structure near the GC observed in 0.5$-$1.5 keV is related with the time duration starting from the quenching of the past activity of Sgr A*. From run B, we find that the time duration should be no more than $\sim$ 0.2 Myr. Observations also show that we can only see the east (left) X-ray structure of the Northern Sky.
Together with the bending of the northern {\it Fermi} Bubble, we speculate that this phenomenon may be caused by the galactic wind blowing from the east to the west in the Northern Sky. Hence, the forward shock in the east will be stronger than the west, and both the temperature and the density of the shocked ISM in the east will be larger than the west, inducing the asymmetric structure of X-ray emission. Another possibility is that the initial ISM is not symmetric, with the density in the left (east) part being higher.

The {\it ROSAT} X-ray structure looks like  an X-ray cavity. Such kinds of cavities have also been observed in other galaxies or galaxy clusters. Usually people think they are formed by the interaction between jets and ISM or intergalactic medium.
However our result reminds us that these cavities may be well formed by the interaction between the winds (rather than jet!) from the central AGN and the IGM. As pointed out by \citet{Young2002} and \citet{DiMatteo2003} in the case of the cavity in M87, if the cavity were formed by the jet, we would expect a sharp bow shock regions between a jet and surrounding medium. This structure  has never been observed.

\subsection{The Effects of Viscosity and Thermal Conduction}

The morphology also depends on the viscosity coefficient $\mu$, as shown by  Fig. \ref{label4}. We can see from the figure that if the viscosity coefficient is larger, the bubble will be more spherical. The winds near the CMZ  suffer from the viscous force because of the large velocity gradient on the $X$$-$$Y$ plane, and they are slowed down by the CMZ significantly. If viscosity coefficient is larger, the kinetic energy of wind gas will be dissipated into internal energy more efficiently, then the thermal pressure close to the GC will be
larger. Therefore, the opening angle of the blown-up CMZ gas will be wider, which causes the bubble more spherical.

Viscosity can suppress both the Rayleigh-Taylor (RT) and Kelvin-Helmholtz (KH) instabilities. Following equations (18) and (19) from \citet{Yang2012}, we can estimate the timescales for the growth of both instabilities.
For example, for the RT instability to form a $\sim 1$ kpc structure at height $z=4$ kpc, the required timescale is about 5 Myr; while for the KH instability, it is about 1.5 Myr. So both instabilities can grow up during the formation of the {\it Fermi} Bubbles. From the left plot of Fig. \ref{label4} we can see that, when viscosity is not included, large rolls with typical length scales of $\sim$ kpc inside the shocked ISM are formed. However, for all the other runs with viscosity included, no such rolls are found.

Another role of viscosity is the viscous heating. This effect is important in the interaction region between winds and CMZ. In this region, the main components of viscous stress tensor are $T_{xz}$ $(=T_{zx})$ and $T_{yz}$ $(=T_{zy})$,
  \be
  \frac{\partial e}{\partial t} \sim T_{xz} \frac{\partial v_{x}}{\partial
  z}+T_{yz} \frac{\partial v_{y}}{\partial z}.
  \ee
For the interaction region, $T_{xz} \simeq \mu {\partial v_{x}}/{\partial z}$. Replacing $\partial v_x$ with $0.1\%c$ and $\partial z $ with $100$ pc, we can estimate that the time-scale for the winds to pass through this  region is $\sim$ 1 Myr. The density of the  CMZ gas blown up in the interaction region is $\sim 10^{-2}~{\rm cm}^{-3}$.  Then the increase of temperature is $\sim 10^{6}$ K.

What is the role of thermal conduction? We study this problem in run E (without thermal conduction). Fig. \ref{label6} shows the distribution of the electron number density and temperature. A jet-like structure along the $z$-axis is clearly seen.
Since the grids along the $z$-axis are elongated, we need to check whether this feature is artificial or not. We have done such a test and found that this feature is likely real. The temperature of this structure is relatively low but the density is high. Their formation mechanism is as follows. In the inner region, the massive CMZ gas acts like a wall around Sgr A*, preventing the winds from expanding in the horizontal direction. The winds collide with the CMZ and the kinetic energy of winds is converted into thermal energy, thus the temperature and pressure increase. The high-pressure gas then escapes towards the polar direction, squeezing the wind from Sgr A* and causing the formation of this jet-like structure.
However, when we include thermal conduction as in most of our runs, this structure  disappears. This is because  thermal conduction can efficiently transport energy between the regions with different temperature thus smooth out this structure.

\subsection{The Role of ISM Density}

\begin{figure*} 
 \includegraphics[width=0.41\textwidth]{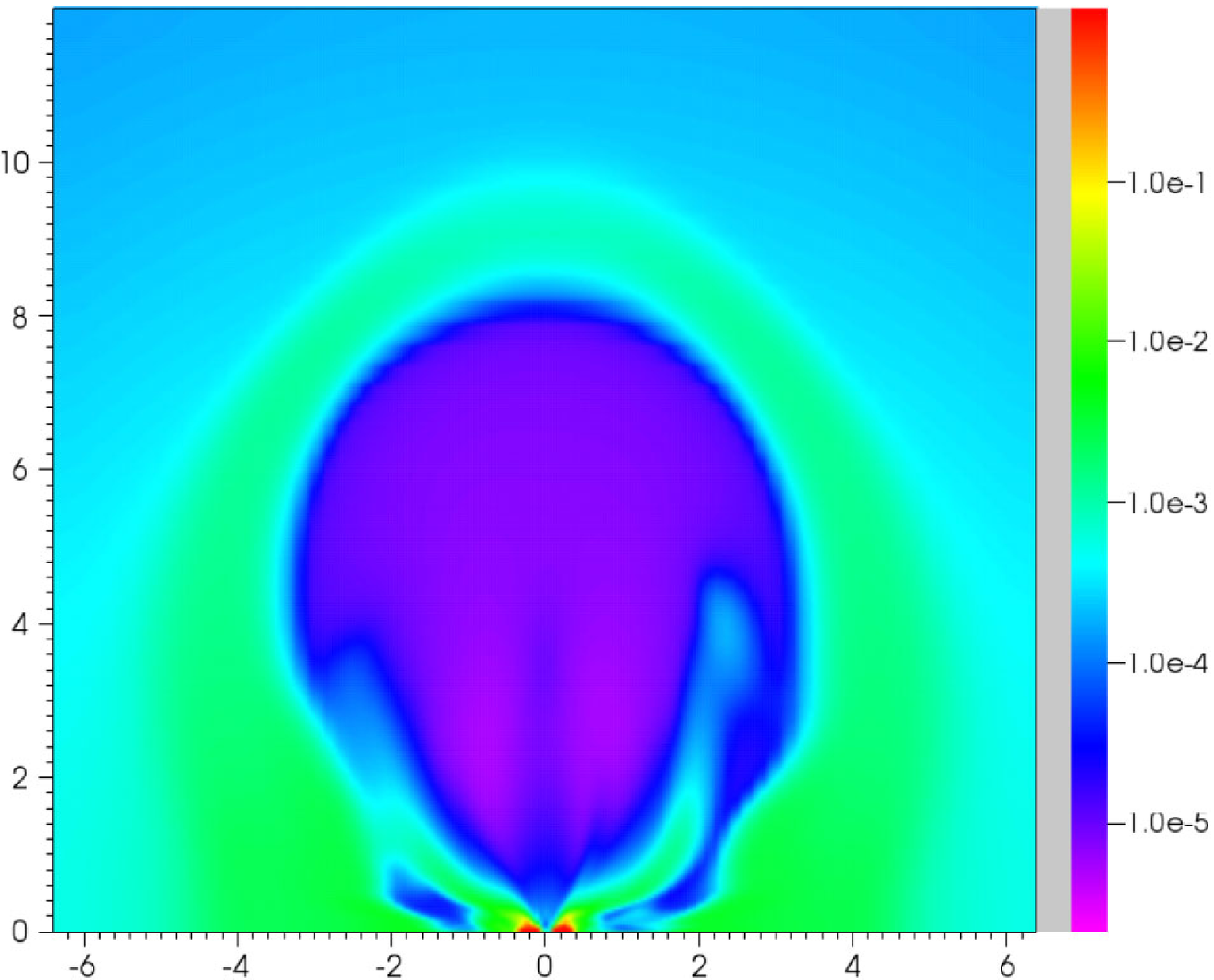}\hspace{2.0cm}
 \includegraphics[width=0.41\textwidth]{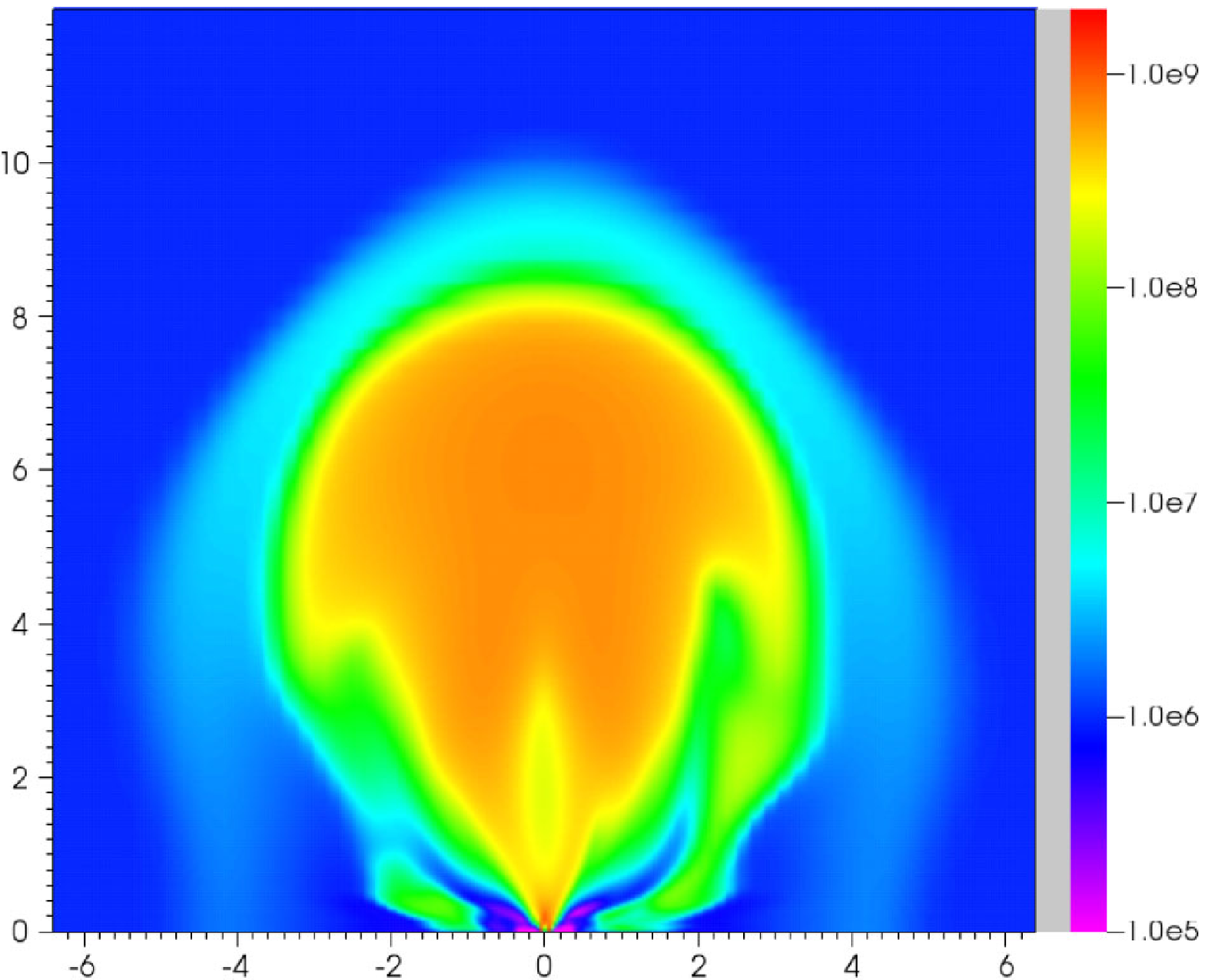}\hspace{1.2cm}
 \caption{The distributions of number density (left) and temperature (right) of electrons on $Y=0$ plane for run H (with a larger transition radius $R_{\rm tr}$ than run A) at $t=11.6$ Myr. }
\label{label9}
\end{figure*}

For simplification, we have adopted a power-law distribution for the density distribution of the initial ISM: $\rho=A/r^{n}$, where \emph{A} and \emph{n} are constants, and $r$ is in unit of $20$ pc in our simulations. The value of \emph{A} has a weak influence on the age of the {\it Fermi} bubbles. This can be seen in run F (refer to Fig. \ref{label7}). Although the density is two times higher than the basic run, it only takes 20\% more time to form the bubbles. This is easy to understand from equations (\ref{Equ_R2}) and (\ref{Equ_Rfan}).

Different from the parameter \emph{A}, the value of \emph{n} is more important to influence the evolution of the bubbles. For example, our simulations show that if the index \emph{n} is changed into 2.0 from 1.6 while \emph{A} remains unchanged compared to run F, the age of the bubble would be 7 Myr, which is half of the age of run F. The temperature of the shocked ISM in latitude $\ga +40^{\circ}$ is about 1 keV in this case.
Physically, this is because the kinetic energy of shocked ISM is nearly a constant fraction of the total energy injected from GC. Specially, from equations (\ref{Equ_R2}) and (\ref{Equ_Rfan}), we can obtain:
\be
\frac{1}{2}M_{c}\dot{R}^{2}_{2}=2\pi A \frac{f^{5-n}}{3-n}P_{w}t,\label{energyequ}
\ee
where $M_{c}$ and $\dot{R}_{2}$ are the mass and velocity of shocked ISM respectively.
We find that $2\pi A \frac{f^{5-n}}{3-n}$ is almost a constant: $\sim 0.3$. So we can approximately rewrite the righthand side of equation (\ref{energyequ}) as: $\eta P_{w}t=\frac{1}{2}M_{c}\dot{R}^{2}_{2}\sim \frac{1}{2}M_{c}R^{2}_{2}/t^{2}$ ($\eta$ is a constant), or $t\propto M^{1/3}_{c}$ when $P_{w}$ keeps unchanged.
For the two cases mentioned above, if \emph{A} is doubled, $M_{c}$ is doubled, while \emph{n} changed from 1.6 to 2.0, $M_{c}$ is only 1/8 of the former. That's why the ages are 20\% larger and one-half smaller respectively.

\subsection{The Role of the Wind Parameters}

In run G, the mass flux of winds is three times higher than run A, while the wind velocity is the same. The age is 34\% shorter than run A. This is very close to the result of 31\% by the simple analytical analysis shown by equation (\ref{Equ_R2}). Because of the increase of the wind power, the velocity of the forward shock increases thus the temperature of the shocked ISM becomes higher, while the increase of temperature inside the bubble is not so obvious.

In run H, we reduce the wind velocity but keep the kinetic power of wind ($P_{w}$) unchanged compared to run A. We find that the kinetic and thermal energy and temperature of different regions do not change much (refer to Table \ref{table2}). This means that, the results mainly depend on the kinetic power, while the velocity and mass outflow rate are degenerate. This is also easy to be understood from equations (\ref{Equ_R2}) and (\ref{Equ_P}).

The gas inside the bubbles mainly comes from the blown-up CMZ gas instead of winds injected from Sgr A*. In addition, the evaporation of the bubble edge, which is determined by thermal conduction, should also supply additional gas. But since there is large uncertainty on thermal conduction coefficient,  it is hard for our model to predict the exact mass of the gas inside the bubbles. In addition to the total mass, another interesting quantity is the temperature of the gas inside the bubbles. To estimate the temperature, we need to know the pressure. This quantity is equal to the gas pressure of the shocked ISM, which is well determined by the density of ISM and the wind power from equation (10). Unfortunately, the uncertainty of the gas density inside the bubbles mentioned-above makes it hard to precisely predict the temperature. Based on our simulation of run A, we can only estimate the temperature of gas within the bubbles to be in the range of $10^{8}\sim 10^{9}$ K. Observational constraints are still lacking.

\section{SUMMARY AND DISCUSSION}

We have performed hydrodynamical numerical simulations to study the formation mechanism of the {\it Fermi} bubbles detected by {\it Fermi}-LAT. Our main aim in the present paper is to explain the morphology and the thermodynamical properties of the bubble, but leaving the study of the production of $\gamma$-ray photons and the explanation of the spectrum to our next work. While Sgr A* is quite dim at the present stage, many observational evidences indicate that this source should be much more active in the past. Specifically, one possibility suggested by a previous work is that the mass accretion rate of the hot accretion flow in Sgr A* should be $10^3-10^4$ times higher than the present value and this activity lasts for several Myr (Totani 2006). Based on this scenario, we show that the observed {\it Fermi} bubbles can be well formed by the interaction between the winds launched from the ``past'' hot accretion flow and ISM. In our model, the winds last for $10^7$ yr and the activity of Sgr A* was quenched no more than $0.2$ Myr ago.
The properties of wind such as the mass flux and velocity are not free parameters but obtained from the previous works on MHD numerical simulations of hot accretion flows. Viscosity and thermal conduction are included which can suppress various instabilities and make the gas inside the bubble uniform. The required power of the winds is $\sim 2\times 10^{41}~\ergs$, which is fully consistent with the previous studies on the past activity of Sgr A*.
The edge of the bubbles corresponds to the contact discontinuity which is the boundary between the shocked interstellar medium and the shocked winds. Properties of the bubbles such as the morphology and the total energy are consistent with observations. The limb-brightened {\it ROSAT} X-ray structure can be interpreted by the shocked ISM behind the forward shock, while the conical-like X-ray structure close to Galactic center is interpreted by the interaction region of wind gas and CMZ gas. Our model can also quantitatively explain both the thermal pressure and the temperature of the X-ray structure in high latitude position ($\ga +40^{\circ}$) revealed by the recent {\it Suzaku} observations.

In addition to winds, jets should also co-exist with hot accretion flow (Yuan \& Narayan 2014). In our model, we do not include the jet. We assume that the interaction between jet and the interstellar medium is negligible because, by definition, jet must be well-collimated and be as fast as the light. In this case, we expect that the jet will simply drill through the ISM, with almost no interaction with the ISM in the Galaxy.

We have also calculated the energy transformation efficiency in our model. We find that at $r\sim10$ kpc, $\sim$ 60\% of the total energy of winds injected from Sgr A* is transported into the ISM. Obviously, such a high efficiency is because of the large opening angle of winds. This result suggests that we may consider the role of winds in solving the cooling flow problem in some elliptical galaxies and galaxy clusters.
Usually people consider the heating of ISM or intracluster medium by jets (see, e.g., \citealt{Vernaleo2006} and references therein). However, numerical simulations have found that jet may only be able to deposit their energy at $r>100$ kpc thus not very efficient (\citealt{Vernaleo2006}). Some solutions have been suggested, e.g., the precession of a jet, or motions of intracluster medium (see \citealt{Vernaleo2006} and \citealt{Heinz2006}).
But another possible way is to invoke winds whose existence has been firmly established by both observational and theoretical studies. Given our successful explanation of the formation of the {\it Fermi} bubbles by the wind model, it is also worthwhile to study whether the X-ray cavities observed in galaxy clusters (e.g., \citealt{Fabian2012}), which have the similar morphology with the {\it Fermi} bubbles, can be produced by winds.

\acknowledgements
We thank Fuguo Xie, Zhaoming Gan, Tomohisa Kawashima, Chao Liu, Hui Li, and Daniel Wang for helpful discussions. We also thank the referee for constructive suggestions.
This work was supported in part by the Natural Science Foundation of China (grants 11103061, 11133005, 11121062, and 11103059), the National Basic Research Program of China (973 Program, grant 2014CB845800), and the Strategic Priority Research Program ``The Emergence of Cosmological Structures'' of the Chinese Academy of Sciences (grant XDB09000000). M. Y. Sun acknowledges support from the China Scholarship Council (No. [2013]3009). Support for the work of M.S. was provided by NASA through Einstein Postdoctoral Fellowship grant number PF2-130102 awarded by the Chandra X-ray Center, which is operated by the Smithsonian Astrophysical Observatory for NASA under contract NAS8-03060.
{The simulations were carried out at SHAO Super Computing Platform.}

\end{document}